\documentstyle[preprint,epsbox,eqsecnum,prd,aps]{revtex}

\newcommand{\beq}{\begin{equation}}
\newcommand{\eeq}{\end{equation}}
\newcommand{\beqa}{\begin{eqnarray}}
\newcommand{\eeqa}{\end{eqnarray}}
\newcommand{\bea}{\begin{array}}
\newcommand{\ena}{\end{array}}

\begin{document}
\title{Generality of  Singularity Avoidance in Superstring Theory:
Anisotropic Case}
\author{Hiroki  Yajima$^1$\thanks{electronic
address:yajima@gravity.phys.waseda.ac.jp},
~Kei-ichi Maeda$^{1,2}$\thanks{electronic
address:maeda@gravity.phys.waseda.ac.jp}, and
Hidetoshi  Ohkubo$^3$\thanks{electronic
address:ohkubo-h@post.yamaha.co.jp
}\\~}
\address{$^1$ Department of Physics, Waseda University,
Shinjuku, Tokyo 169-8555, Japan\\[.3em]
$^2$ Advanced Research Institute for Science and Engineering,\\
Waseda University, Shinjuku, Tokyo 169-8555, Japan\\[.3em]
$^3$ Engineering Department,
 YAMAHA CORPORATION,  Shizuoka 438-0192, Japan\\~}
\maketitle
\begin{abstract}
\baselineskip 18pt
In one-loop string effective action, we study a generality of non-singular cosmological solutions  found in the isotropic and homogeneous case. We discuss Bianchi I and IX type spacetimes. We find that nonsingular solutions still exist in Bianchi I model around nonsingular flat Friedmann  solutions. On the other hand, we cannot find any nonsingular solutions in Bianchi IX  model. Non-existence of nonsingular Bianchi IX universe may be consistent with the analysis by Kawai, Sakagami and Soda, i.e. the tensor mode perturbations against nonsingular flat Friedmann universe are unstable, because Bianchi IX model can be regarded as a closed Friedmann universe with a single gravitational wave. So based on these facts, we may conclude the nonsingular universe is found in isotropic case is generally unstable, a singularity avoidance may not work in the present model.
\end{abstract}

%
\section{Introduction}
%

\ The initial singularity is one of the most
serious problems in the Big Bang universe.
Even the inflationary universe model, which  resolves many
difficulties in the early universe,  cannot provide us an avoidance of
initial singularity.
Quantum effect might resolve it as a scenario in quantum cosmology, but
quantum gravity is not yet completed.
To construct a theory of quantum gravity, a unification of fundamental
interactions may be one of the most promising ways.  Among many
attempts of such a unification, a superstring model would be the best
candidate\cite{witten1}.  In fact,
the origin of a black hole entropy may be understood by a string
theory.
Hence, the initial singularity problem might also be solved in the
context of a string theory, which may provide us to  manage the physics at Planck scale. However a full theory have not yet developed.
We may so far not be able to discuss the early universe in a
string theory as it is, except for some restricted models\cite{??}.
Therefore many efforts to find
nonsingular universe are mainly based on a low energy effective field
theory of a superstring model.
Such an effective field theory will
break down beyond the Planck energy scale. However, if a superstring
theory is truly the theory of everything, its effective theory may reveal
some important aspects about gravity.
In particular, if it has a property of singularity avoidance, it would be
a nice  evidence for a
superstring.

Based on an effective theory, there is one interesting approach
called  pre-big-bang universe model\cite{gasp1}. This model is based on
T-duality, which gives some relation  between large and small scales.
Then, assuming that  a  scale factor of the Universe  has such a
duality\cite{vene1}, we find a
cosmological solution which consists of two distinct and disconnected
branches. One of the branches($t>0$) corresponds to the
expanding Friedmann universe, and other branch($t<0$) gives another expanding universe, ending with a
singularity at
$t=0$. In pre-big-bang scenario, however,  those two disconnected branches
are assumed to be connected  without a singularity on account of some
unknown stringy  effect, which is not included in the lowest effective
action.

There are several works to study whether two branches can be connected
without a singularity.   At
the tree-level in a superstring action, however, a possibility of
classical branch-changing solutions is
excluded\cite{east2}. Only quantum effect may be a hope to
connect two branches. In fact,  taking into one-loop contributions to a
superstring effective action with dilaton and modulus
fields\cite{anton1}, Antoniadis, Rizos and Tamvakis found an existence of
nonsingular solution with spatially flat background, with the same action, then, Easther and one of the present authors also showed a nonsingular closed universe\cite{east1}. Although those solutions do not help the pre-big-bang scenario, they may be still interesting because of singularity avoidance.

A question about such nonsingular solutions, however, is whether those are generic or not. Although we find nonsingular solutions for some finite range of
initial data, the spacetime is assumed to be isotropic and homogeneous\cite{east1}. The universe, however, may begin with an anisotropic and/or inhomogeneous
geometry.  Therefore, we have to study whether the present
nonsingular solution is generic or not, or is stable or unstable.
Recently, Kawai, Sakagami and Soda analyzed stability of flat nonsingular
solution against perturbations, and found that a tensor mode is unstable.
However, since a closed nonsingular universe bounces in a finite time, it
is not clear whether it is unstable against tensor mode perturbations.
 In order
to clarify such a problem and to study generality of nonsingular
solutions, in this paper, we  will analyze two types of Bianchi models;
Bianchi I and IX model.

This paper is organized as follows.
In the next section, we introduce the basic equations
for Bianchi I and IX models.
The numerical results are  presented in Sec. III, and conclusions
and discussions follow. We adopt the metric signature  $(-,+,+,+)$ and
units of $c = 8 \pi G = 1$.

%
\section{Basic Equations}
%

\ We take the  following one-loop effective
action\cite{anton1} -\cite{rizos1}
\beqa
S = \int
d^{4}x\sqrt{-g}\Bigl[\frac{1}{2}R
-\frac{1}{4}\nabla_{\mu}\phi\nabla^{\mu}\phi
-\frac{3}{4}\nabla_{\mu}\sigma\nabla^{\mu}\sigma
-\frac{1}{6}H^{\mu\nu\lambda}H_{\mu\nu\lambda} +\frac{1}{16}[\lambda
e^{\phi}-\delta \xi(\sigma)]R^{2}_{GB}\nonumber \\ +({\rm
Higher\,Curvature\,Terms})\Bigr],
\eeqa
where $R$, $\phi$ and $\sigma$ are the scalar curvature,  the
dilaton, and the modulus field, respectively.
\beqa
R^{2}_{GB}=
R^{2}-4R^{\mu\nu}R_{\mu\nu}
+R^{\mu\nu\alpha\beta}R_{\mu\nu\alpha\beta}
\eeqa
 is the
Gauss-Bonnet term, and
$H$ is the antisymmetric tensor field. The
coefficient $\lambda$ is positive definite and determined by the inverse
string tension $\alpha'$ and string coupling $g$. The
coefficient $\delta$ depends on the relative numbers of chiral,
vector and spin-$\frac{3}{2}$ massless supermultiplets and is
proportional to the four-dimensional trace anomaly of the $N = 2$
sector. The $\delta$ would be either  positive
or negative. The  function $\xi(\sigma)$ is
defined as
\beqa
\xi(\sigma)=\ln[2 e^{\sigma} \eta^{4}(i e^{\sigma})]
\eeqa
by the Dedekind $\eta$ function\cite{erd}
\beqa
\eta(\tau)=q^{\frac{1}{12}} \prod_{n=1}^{\infty}(1-q^{2n}),
~~q
=e^{i\pi\tau}.
\eeqa
The first derivative of $\xi$ with respect to $\sigma$ is
\beqa
\xi_{\sigma}(\sigma)=1-\frac{\pi e^{\sigma}}{3}+8\pi
e^{\sigma}\sum_{n=1}^{\infty}\frac{n e^{-2n \pi e^{\sigma}}}{1-e^{-2n
\pi e^{\sigma}}},
\label{eqn:exact}
\eeqa
which can be approximated by $\sinh  \sigma$ as
\beqa
\xi_{\sigma}(\sigma)\approx
-\frac{2\pi}{3}\sinh \sigma
\label{eqn:approx}.
\eeqa
As was shown in \cite{east1}, Eq. (\ref{eqn:approx}) is
very close to the exact function
(\ref{eqn:exact}). Then we will use  Eq.(\ref{eqn:approx}) in our
analysis just for simplicity, although we have also checked our results
by use of  the exact function.
We also introduce the following function $f(\phi,\sigma)$ for
convenience:
\beqa
f(\phi,\sigma)=\frac{1}{16}[e^{\phi}-\bar{\delta} \xi(\sigma)],
\eeqa
where $\bar{\delta}\equiv\delta/\lambda$.
We  set $H \equiv 0$ and ignore  higher curvature terms
than second order.

With the same model, cosmological solutions have been
studied from a view point of the initial singularity problem.
Antoniadis, Rizos and
Tamvakis analyzed a spatially flat Friedmann model and found nonsingular
solution\cite{anton1}.
Easther and Maeda extended their analysis to a closed Friedmann
model and also showed nonsingular solution\cite{east1}. Since we wish to
examine  generality of those nonsingular solutions, we
extend their works to anisotropic spacetimes. Here we study only  Bianchi
I and IX models because those spacetimes include a flat and a closed
Friedmann models.

It maybe convenient to rescale time and spatial coordinates by $\lambda$ as $\bar{t}=t/\sqrt{\lambda}$, $\bar{x^{i}}=x^{i}/\sqrt{\lambda}$. Hereafter
we drop a bar for convenience.
Taking a variation of the action,  we obtain  the basic equations. With
those basic equations, for Bianchi I and IX models, we can assume the
following diagonal metric form:
\beqa
ds^{2}=-dt^{2}+e^{-2 \Omega}e^{2 \beta_{ij}}\omega^{i}\omega^{j},
\eeqa
\[
\beta_{ij} = \left[
\begin{array}{ccc}
\beta_{+}+\sqrt{3} \beta_{-}  & 0  & 0  \\
0  & \beta_{+}-\sqrt{3} \beta_{-}  & 0  \\
0  & 0  & -2 \beta_{+}
\end{array}
\right]
\]
In both cases, we also introduce scale factors $a(t)$,$b(t)$, and $c(t)$
as
\beqa
e^{p(t)}=e^{- \Omega +\beta_{+}+\sqrt{3} \beta_{-}},~~
e^{q(t)}=e^{- \Omega + \beta_{+}-\sqrt{3} \beta_{-}},~~
e^{r(t)}=e^{- \Omega -2\beta_{+}}.
\eeqa

The basic equations obtained are divided into two groups:\\
(1) The dynamical equations for the metric $p, q$ and $r$, the dilaton
field $\phi$, and the modulus field
$\sigma$ :
\beqa
&& (1+8\dot{r}\dot{f})(\ddot{q}+\dot{q}^{2})
+(1+8\dot{q}\dot{f})(\ddot{r}+\dot{r}^{2})
+(1+8\ddot{f})\left[\dot{q}\dot{r} + {1 \over 2}
U_1\right]+\frac{1}{4}\dot{\phi}^{2}
+\frac{3}{4}\dot{\sigma}^{2}=0,\label{eq_p}
\\ && (1+8\dot{r}\dot{f})(\ddot{p}+\dot{p}^{2})
+(1+8\dot{p}\dot{f})(\ddot{r}+\dot{r}^{2})
+(1+8\ddot{f})\left[\dot{r}\dot{p} + U_2\right]
+\frac{1}{4}\dot{\phi}^{2}+\frac{3}{4}\dot{\sigma}^{2}=0, \label{eq_q} \\
&& (1+8\dot{q}\dot{f})(\ddot{p}+\dot{p}^{2})+(1+8\dot{p}\dot{f})(\ddot{q}
+\dot{q}^{2})+(1+8\ddot{f})\left[\dot{p}\dot{q}
+U_3\right]+\frac{1}{4}\dot{\phi}^{2} +\frac{3}{4}\dot{\sigma}^{2}=0,
\label{eq_r} \\
&& \ddot{\phi}+(\dot{p}+\dot{q}+\dot{r})\dot{\phi}
=2\frac{\partial f}{\partial\phi}R^{2}_{GB},
\label{eq_phi} \\
&& \ddot{\sigma}+(\dot{p}+\dot{q}+\dot{r})\dot{\sigma}
=\frac{2}{3}\frac{\partial f}{\partial\sigma}R^{2}_{GB},
\label{eq_sigma}\eeqa
where an overdots denotes a differentiation with respect to $t$, and the
Gauss-Bonnet term $R^{2}_{GB}$ is given as
\beqa
R^{2}_{GB}&=&8 \Bigl[\bigl[\ddot{p}
+\dot{p}^{2}\bigr]\bigl[\dot{q}\dot{r}+\frac{1}{2}U_1\bigr]+\bigl[\ddot{q}
+\dot{q}^{2}\bigr]\bigl[\dot{r}\dot{p}+\frac{1}{2}U_2\bigr]+\bigl[\ddot{r}
+\dot{r}^{2}\bigr]\bigl[\dot{p}\dot{q}+\frac{1}{2}U_3\bigr]
\nonumber\\
& &
+ {1 \over 2}\dot{p}^{2}{\partial U_1 \over \partial p}
+ {1 \over 2}\dot{q}^{2}{\partial U_2 \over \partial
q}
+ {1 \over 2}\dot{r}^{2}{\partial U_3 \over \partial r}
-{1 \over
2}\dot{p}\dot{q}[{\partial U_1 \over \partial p}+{\partial U_2 \over
\partial q}-{\partial U_3 \over \partial r}]
\nonumber\\
& &
-{1 \over
2}\dot{q}\dot{r}[{\partial U_2 \over \partial q}+{\partial U_3 \over
\partial r}-{\partial U_1 \over \partial p}]
-{1 \over
2}\dot{r}\dot{p}[{\partial U_3 \over \partial r}+{\partial U_1 \over
\partial p}-{\partial U_2 \over \partial q}]\Bigr].
\eeqa
(2) The constraint equation :\\
\beqa
& & \dot{p}\dot{q}+\dot{q}\dot{r}+\dot{r}\dot{p}
+24\dot{p}\dot{q}\dot{r}\dot{f}+4\dot{p}\dot{f}U_1
+4\dot{q}\dot{f}U_2
+4\dot{r}\dot{f}U_3 -\frac{1}{4}\dot{\phi}^{2}
-\frac{3}{4}\dot{\sigma}^{2}
\nonumber\\
& &
~~~~~~
+\frac{1}{2}(U_1 +U_2+U_3)
=0 .
\label{constraint}\eeqa
Here the functions $U_1, U_1, U_1$ are  defined by \\
(A) Bianchi I model :\\
\beqa
U_1=U_2=U_3=0
\eeqa
(B) Bianchi IX model :\\
\beqa
U_1 &=& e^{-2q}+e^{-2r}-e^{-2p}
+\frac{1}{2}(e^{2(q-r-p)}+e^{2(r-p-q)}
-3e^{2(p-q-r)})\nonumber\\
U_2 &=& e^{-2r}+e^{-2p}-e^{-2q}
+\frac{1}{2}(e^{2(r-p-q)}+e^{2(p-q-r)}-3e^{2(q-r-p)})\nonumber\\
U_3 &=& e^{-2p}+e^{-2q}-e^{-2r}
+\frac{1}{2}(e^{2(p-q-r)}+e^{2(q-r-p)}-3e^{2(r-p-q)}).
\eeqa

The basic equations above are the five second-order differential
equations with one constraint equation. For  $p(t) = q(t) = r(t)$, i.e.
the case of isotropic and homogeneous model,  the equations
are  reduced to the cases of a flat universe studied by Antoniadis, Rizos
and
Tamvakis\cite{anton1}, and of a close universe by Easther and
Maeda\cite{east1}.

\section{Numerical Results}
\ We have examined the case of $\delta < 0$  because in the isotropic and
homogeneous case, nonsingular cosmological solutions are
found  only for
$\delta < 0$\cite{anton1},\cite{east1}.
We solve the basic equations numerically.

Since five second-order derivatives
$\ddot{p},\ddot{q},\ddot{r},\ddot{\phi},\ddot{\sigma}$ in the basic
equations (\ref{eq_p})-(\ref{eq_sigma}) are coupled,
we have to make an inverse transformation as follows:
Defining a vector
$\mbox{\boldmath $x$}=(\ddot{p},
\ddot{q},
\ddot{r},
\ddot{\phi},
\ddot{\sigma}
)$, the basic equations  (\ref{eq_p})-(\ref{eq_sigma}) are written in the
matrix form as
\beqa
Z \mbox{\boldmath $x$} = \mbox{\boldmath $y$},
\label{basic_eq}
\eeqa
where
$5\times5$ matrix $Z=Z(p,q,r,
\phi,\sigma,\dot{p},\dot{q},\dot{r},\dot{\phi},\dot{\sigma})$
and vector $\mbox{\boldmath
$y$}=\mbox{\boldmath $y$}(p,q,r,
\phi,\sigma,\dot{p},\dot{q},\dot{r},\dot{\phi},\dot{\sigma})$ are
known explicitly from the basic equations.

Unless $\Delta \equiv detZ$ vanishes, we have
\beqa
\mbox{\boldmath $x$} = Z^{-1} \mbox{\boldmath $y$},
\label{basic_eq2}
\eeqa
then
we can solve the basic equations (\ref{basic_eq2}), giving
initial data of $p, q, r,
\phi,\sigma,\dot{p},\dot{q},\dot{r},\dot{\phi},\dot{\sigma}$.
 However, when $\Delta$ vanishes,
then we cannot proceed our numerical calculations further. Such a
end point, which may appear in the evolution of the Universe,
 seems to be a spacetime
singularity, however more detail analysis will be required as we will show
later.

Although we have used scale factors $p, q$ and $r$ for the basic
equations, we shall describe our results by  two anisotropy variables
$\beta_{+}$ and
$\beta_{-}$.

\subsection{Bianchi I Type Case}
\ First we will show the results in Bianchi I model.
We choose $\bar{\delta} = -48/\pi$.
We introduce scale factors as
\beqa
a(t) = e^{p(t)}=e^{-\Omega +\beta_++\sqrt{3}\beta_-}, b(t) =
e^{q(t)}=e^{-\Omega +\beta_+-\sqrt{3}\beta_-}, c(t) = e^{r(t)}=e^{-\Omega
-2\beta_+}.
\eeqa
Without loss of generality, we can set $\Omega_{0}=\beta_{\pm}=0$, i.e.
$a_0=b_0=c_0=1$. Here the subscript 0 denotes an initial value of the
variables. We define the initial time $t=t_{0}=0$.
We have to give initial data of $\dot{\Omega}_0, \dot{\beta_{\pm}}_{0},
\phi_{0}, \dot{\phi}_{0}, \sigma_{0}$ and $\dot{\sigma}_{0}$, which must satisfy one constraint equation (\ref{constraint}).  Then we have five independent
initial values.  Since we are interested in whether nonsingular solutions found in the isotropic case are generic or not, we shall set up the initial data in the
anisotropic case around the isotropic ones.

In the isotropic and homogeneous case, we find nonsingular solutions for some
finite parameter range of initial data. In Fig. 1, fixing $\dot{\Omega}_0 =-0.1, \sigma_0=0$ we show the range of initial data of $\phi_{0}$ and $\dot{\phi}_{0}$, which gives for nonsingular solutions (shown by a circle).  $\dot{\sigma}_0$ is determined by the constraint equation (\ref{constraint}). In Fig. 1, if $\dot{\phi}_{0}$ is efficiently small, then in the case of $\phi_{0}<0$ there always exist nonsingular solutions. We think because if $\phi<0$ and $|\phi|\gg 1$, then $e^{\phi}\approx 0$ so singularity avoidance is almost independent of $\phi$.

To give initial data $\dot{\Omega}_{0}, \phi_{0}, \dot{\phi}_{0}$ and $\sigma_{0}$ with an anisotropy, taking one nonsingular isotropic solution, including anisotropy, i.e. $\dot{\beta_{+}}_{0}, \dot{\beta_{-}}_{0} (\neq 0)$, and solving for $\dot{\sigma}_{0}$ by the constraint equation we have set up our initial data.

We find nonsingular solutions in Bianchi I model near the isotropic nonsingular solution. We show  one example in Fig.2, where we have set  $\dot{\Omega}_{0}=-0.1$, $\phi_{0}=\dot{\phi}_{0} = 0$, $\sigma_{0} = 0$ and $\dot{\beta_{+}}_{0} = 0.05, \dot{\beta_{-}}_{0}=0.0$, $\dot{\sigma}_{0}=0.17320508075689$, which has been determined by the constraint equation ($\dot{\sigma}_{0}=0.200$ for the isotropic case). We also show a singular solution in Fig. 3. The volume $e^{\Omega}$ vanishes at $t \approx -4.98$ and $t \approx 8.35$, resulting in a Big Bang type singularity.

Changing the anisotropic parameters $\dot{\beta_{+}}_{0}, \dot{\beta_{-}}_{0}$, we search for nonsingular solution in order to find how large anisotropy is possible to allow nonsingular solutions.
The result is shown in Fig. 4 for $\dot{\Omega}_{0}=-0.1$, $\phi_{0}=\dot{\phi}_{0}=0$, $\sigma_{0}=0$.  We find that if anisotropy is enough large at the initial stage, then the spacetime evolves into a singularity.
The boundary between non-singular and singular solutions in $(\dot{\beta_{+}}_{0},\dot{\beta_{+}}_{0})$-plane is almost a circle. $\Delta r/r \leq O(10^{-2})$.
Whether spacetime will evolve into a singularity or avoid it, of course, strongly depends  on initial parameters which we have set ($\dot{\Omega}_{0}, \phi_{0}, \dot{\phi}_{0}, \sigma_{0}$).
As we move these initial parameters to the boundary values in the isotropic case beyond which no non-singular solution is found, the radius of the boundary in $(\dot{\beta_{+}}_{0},\dot{\beta_{+}}_{0})$-plane decreases and if we set initial parameters fairly close to the boundary values, then the range for non-singular solution eventually disappears. Then in this case the singularity avoidance does no longer work.

We can conclude that for the Bianchi I type anisotropy, a singularity avoidance is still generic. Because the range of non-singular solutions is finite and not small.

However, we have the stability analysis done by Kawai, Sakagami and Soda\cite{kawai1}\cite{soda1}\cite{kawai3}. Their analysis is based on the same action (but only a dilaton field) and flat Friedmann background case. They found there exists instability in the tensor mode. Then in the case of the flat universe the universe is unstable. This result may not change even in Bianchi I background universe.

\subsection{Bianchi IX Type Case}
\ In the same way in the Bianchi I type case, we will analyze Bianchi type IX. We introduce scale factors as
\beqa
a(t) = 2 e^{p(t)}, b(t) = 2 e^{q(t)}, c = 2 e^{r(t)}.
\eeqa
The factor 2 is definitions of invariant basis and the scale factor in a closed Friedmann model. As the same as Bianchi I model, we first research for closed nonsingular Friedmann solutions. We search for non-singular Bianchi IX solutions.

First setting up the initial anisotropic data around those in a closed nonsingular solution, we show an isotropic non-singular solution in Fig. 5. We choose a negative value of $\bar{\delta}$ as $\bar{\delta}=-48/\pi$, and set $a(0) = b(0) = c(0) = e^{3.2514}$, $\dot{\Omega}_{0}=-0.01$, $\phi_{0}=-3$, $\dot{\phi}_{0}=-5\times 10^{-4}$, $\sigma_{0}=0$ which was found by Easter and Maeda. We can reproduce one of their solutions.

An anisotropy is induced just as the same as Bianchi I model, we show our numerical results in Fig. 6 and Fig. 7. As for anisotropy we set $\dot{\beta_{+}}_{0} = 0.01$ and $\dot{\beta_{-}}_{0} = 0.0$ in Fig.5, and $\dot{\beta_{+}}_{0} = 10^{-14}$ and $\dot{\beta_{-}}_{0} = 0$ in Fig.6, respectively.

In the evolution of the universe, $\Delta$($= det Z$) eventually vanishes. We always find a singularity at a certain finite time, so our numerical analysis have to break down. This type of breakdown appears even if an initial anisotropy is quite small, that is, both $\dot{\beta_{+}}_{0}$ and $\dot{\beta_{-}}_{0}$ are less than $10^{-14}$. As shown in Fig. 6 and Fig. 7, $I = R^{\mu\nu\alpha\beta}R_{\mu\nu\alpha\beta}$ suddenly diverges near the breakdown point.

We show the time evolution of $\beta_{+}$ and $\dot{\beta_{+}}$ in Fig. 8 and Fig. 9, where $\dot{\beta_{+}}_{0} = 0.01$, $\dot{\beta_{-}}_{0} = 0.0$ and $\dot{\beta_{+}}_{0} =10^{-14}$, $\dot{\beta_{-}}_{0} = 0$, respectively. $\beta_{+}$, $\beta_{-}$ or both will also increase very rapidly near the breakdown point. Because of the exponential growth of $I = R^{\mu\nu\alpha\beta}R_{\mu\nu\alpha\beta}$ and $\beta_{\pm}$, we believe it is a curvature singularity. Since the volume factor $e^{\Omega}$ does not vanish there, it is not a Big Bang type singularity. The Universe find a singularity with a finite volume.

We have searched the wide range of parameters where Easther and Maeda found nonsingular closed universe solutions, but we could not find any nonsingular solution even if the anisotropy is extremely small. In all cases we examined $I = R^{\mu\nu\alpha\beta}R_{\mu\nu\alpha\beta}$ and $\beta_{\pm}$ show a sharp increase near the point where $\Delta$ vanishes.
Then we would conclude there is no nonsingular solution in anisotropic Bianchi IX type models and all solutions except for exact isotropic case eventually evolve into a curvature singularity.

Our results may be consistent with the stability analysis by Kawai, Sakagami and Soda\cite{kawai1}\cite{soda1}\cite{kawai3}. King showed that Bianchi IX type model can be regarded as a closed Friedmann background with a single gravitational wave of a fixed wave number $k = \sqrt{6}/S$ where $S$ is the radius of three-sphere\cite{king}. This means that Bianchi IX model is the same as the closed Universe with a nonlinear tensor perturbation. Then, if the tensor mode is unstable even for a closed universe we can understand our results. In fact since we know the wave number $k$, we can estimate the timescale of the tensor mode instability from their analysis and compare it with our numerical results. We find that our breakdown time is the same order of instability timescale, i.e. the order of our breakdown timescale is almost $(0.1 \sim 1) \sqrt{|\delta|}$, which depends on initial parameters, and the instability timescale is $\sim 1 \sqrt{|\delta|}$.

\section{Conclusion Remarks and Discussions}
\ In this paper we have examined generality of previously found cosmological nonsingular solutions in the heterotic superstring effective action in orbifold compactifications with one-loop correction.

In Bianchi I type case, many nonsingular solutions are found around nonsingular flat Friedmann solutions without fine-tuned initial conditions. On the other hand, in Bianchi IX type case, we cannot find any nonsingular solution even if anisotropy is quite small. In this case, anisotropy grows exponentially, and our numerical analysis breaks down at a certain point where anisotropy will diverge. At this point, both $\beta_{\pm}$, which represents the anisotropy, and $I = R^{\mu\nu\alpha\beta}R_{\mu\nu\alpha\beta}$ seems to diverge, then the universe may evolve into a curvature singularity. The volume factor remains a finite value.

Our result is consistent with the stability analysis by Kawai, Sakagami and Soda, who found tensor mode instability in flat Friedmann background.
Because Bianchi IX type model can be regarded as a closed Friedmann background with a single gravitational wave of a fixed wave number. 

Since the nonsingular flat Friedmann model is unstable against tensor perturbations, we may expect that nonsingular Bianchi I model is not generic as well.
Therefore, we may conclude the nonsingular universes found in isotropic cases are not generic and a singularity avoidance may not work in the present model.

We have argued only first order expansion term of inverse string  tension $\alpha'$ and one-loop correction and excluded antisymmetric tensor (a axion field) which might exist in the early universe. For more rigid consideration we may need to study based on a full superstring theory, though such a full theory have not yet been developed into a point where we can deal with cosmology. Then we hope to find some effects of a superstring theory by which $R^{\mu\nu\alpha\beta}R_{\mu\nu\alpha\beta}$ will not diverge in some wide range of initial values.
It might be given by higher curvature terms in the effective action.

\section*{ACKOWLEDGEMENTS}
\ We would like to thank Dmitri V. Gal'tsov, V. Sahni, Jiro Soda, Shinsuke Kawai, Takashi Torii, Jun-ichirou Koga and Motoyuki Saijo for useful discussions.
This work was supported partially by a Grant-in-Aid for   Scientific
Research Fund of the Ministry of Education, Science and Culture
(Specially Promoted Research No. 08102010), and by the Waseda University
Grant
for Special Research Projects.


\newpage
\begin{flushleft}
{ Figure Captions}
\end{flushleft}
\noindent
\parbox[t]{2cm}{FIG. 1:\\~}\ \
\parbox[t]{12cm}
{We show the fate of the flat Friedmann Universe in terms of initial values,  $\phi_{0}$ and $\dot{\phi}_{0}$ in the cases of the flat Universe. $\circ$ means regular solutions, while $\times$ means that the Universe evolve into a singularity. For $\triangle$, the solution seems to be singular, although we could not confirm it because we need more CPU time.
}\\[1em]
\noindent
\parbox[t]{2cm}{FIG. 2:\\~}\ \
\parbox[t]{12cm}
{We show one nonsingular solution. We choose a negative value of $\bar{\delta}$ as $\bar{\delta} = -48/\pi$. 
 We have set  $\dot{\Omega}_{0}=-0.1$,
$\phi_{0}=\dot{\phi}_{0} = 0$, $\sigma_{0} = 0$
and $\dot{\beta_{+}}_{0} = 0.05, \dot{\beta_{-}}_{0}=0.0$.
$\dot{\sigma}_{0}=0.17320508075689$ has been determined by the constraint equation ($\dot{\sigma}_{0}=0.200$ for the isotropic case). We show the scale factor $a$ in $(a)$, dilaton field $\phi$ in $(b)$, modulus field $\sigma$ in $(c)$ and $I = R^{\mu\nu\alpha\beta}R_{\mu\nu\alpha\beta}$ in $(d)$, respectively. The solid line, dashed line and dotted line represent scale factors $a$, $b$ and $c$ in the anisotropic case, respectively. The dash-dotted line represents a scale factor in the isotropic case.
 }\\[1em]
\noindent
\parbox[t]{2cm}{FIG. 3:\\~}\ \
\parbox[t]{12cm}
{We show one singular solution. We choose a negative value of $\bar{\delta}$ as $\bar{\delta} = -48/\pi$. 
 We have set  $\dot{\Omega}_{0}=-0.1$, $\phi_{0}=\dot{\phi}_{0} = 0$, $\sigma_{0} = 0$ and $\dot{\beta_{+}}_{0} = 0.1, \dot{\beta_{-}}_{0}=0.0$.
$\dot{\sigma}_{0}=0.00$ has been determined by the constraint equation ($\dot{\sigma}_{0}=0.200$ for the isotropic case). We show the scale factor $a$ in $(a)$, dilaton field $\phi$ in $(b)$, modulus field $\sigma$ in $(c)$ and $I = R^{\mu\nu\alpha\beta}R_{\mu\nu\alpha\beta}$ in $(d)$, respectively. The solid line, dashed line and dotted line represent scale factors $a$, $b$ and $c$ in the anisotropic case, respectively. The dash-dotted line represents a scale factor in the isotropic case.
}\\[1em]
\noindent
\parbox[t]{2cm}{FIG. 4:\\~}\ \
\parbox[t]{12cm}
{Changing the anisotropic parameters $\dot{\beta_{+}}_{0} and \dot{\beta_{-}}_{0}$, we search nonsingular solutions. For $\bar{\delta} = -48/\pi$, $\dot{\Omega}_{0}=-0.1$ and $\sigma_{0}=0$, we show the boundary on the $\dot{\beta_{+}}_{0}$ and $\dot{\beta_{-}}_{0}$ plane, beyond which no non-singular solution is found. The boundary is almost a circle.
The solid line, dashed line and dotted line represent the boundary in the cases of $\phi_{0}=\dot{\phi}_{0}=0$, of $\phi_{0}=1.5, \dot{\phi}_{0}=0.20$ and of $\phi_{0}=3.0, \dot{\phi}_{0}=0.40$, respectively.
}\\[1em]
\noindent
\parbox[t]{2cm}{Fig. 5:\\~}\ \
\parbox[t]{12cm}
{We show a nonsingular solution in a closed Friedmann universe. We choose a negative value of $\bar{\delta}$ as $\bar{\delta}=-48/\pi$. We set $a_{0} = e^{3.2514}$, $\dot{\Omega}_{0}=-0.01$, $\phi_{0}=-3$, $\dot{\phi}_{0}=-5\times 10^{-4}$, $\sigma_{0}=0$. We show the scale factor $a$ in $(a)$, dilaton field $\phi$ in $(b)$, modulus field $\sigma$ in $(c)$ and $I = R^{\mu\nu\alpha\beta}R_{\mu\nu\alpha\beta}$ in $(d)$, respectively.
}\\[1em]
\noindent
\parbox[t]{2cm}{Fig. 6:\\~}\ \
\parbox[t]{12cm}
{We show a singular solution in Bianchi IX type model. We set the same initial values in Fig. 5, $\bar{\delta}=-48/\pi$, $a_{0} = b_{0} = c_{0} = e^{3.2514}$, $\dot{\Omega}_{0}=-0.01$, $\phi_{0}=-3$, $\dot{\phi}_{0}=-5\times 10^{-4}$, $\sigma_{0}=0$, except for anisotropy, which is chosen as $\dot{\beta_{+}}_{0} = 0.01$ and $\dot{\beta_{-}}_{0} = 0.0$. We show the scale factors $a$, $b$ and $c$ in $(a)$, dilaton field $\phi$ in $(b)$, modulus field $\sigma$ in $(c)$ and $I = R^{\mu\nu\alpha\beta}R_{\mu\nu\alpha\beta}$ in $(d)$, respectively. We find that $I$ grows almost exponentially.
}\\[1em]
\noindent
\parbox[t]{2cm}{Fig. 7:\\~}\ \
\parbox[t]{12cm}
{We show a singular solution in Bianchi IX type model. We set the same initial values as those in Fig.6, except for anisotropy i.e. $\dot{\beta_{+}}_{0} = 10^{-14}$ and $\dot{\beta_{-}}_{0} = 0$. We show scale factors $a$, $b$ and $c$ in $(a)$ and $(b)$, dilaton field $\phi$ in $(c)$, modulus field $\sigma$ in $(d)$ and $I = R^{\mu\nu\alpha\beta}R_{\mu\nu\alpha\beta}$ in $(e)$, respectively. We find that $I$ grows almost exponentially.
}\\[1em]
\noindent
\parbox[t]{2cm}{Fig. 8:\\~}\ \
\parbox[t]{12cm}
{We show the time evolution of $\beta_{+}$ and  $\dot{\beta_{+}}$. We set initial parameters as those in Fig. 6. The anisotropy $\beta_{+}$ also grows exponentially, as $I$ diverges.
}\\[1em]
\noindent
\parbox[t]{2cm}{Fig. 9:\\~}\ \
\parbox[t]{12cm}
{We show the time evolution of $\beta_{+}$ and $\dot{\beta_{+}}$ for the solution given in Fig. 7. The anisotropy $\beta_{+}$ also grows exponentially, as $I$ diverges.
 }\\[1em]
\noindent


\begin{thebibliography}{99}
\bibitem{witten1}
For textbook see e.g. M. B. Green, J. H. Schwarz and E. Witten,  {\it
Superstring theory I, II}, (Cambridge Univ. Press 1987); D. L\"ust
and S Theisen, {\it Lectures on String Theory}, (Springer-Verlag
1989).
\bibitem{??}
To see e.g. A. Lukas, B. A. Overut and D. Waldram, hep-th/9802041; J.D. Lykken, astro-ph/9903026; N. Kaloper, I. I. Kogan and K. A. Olive, Phys. Rev. {\bf D 57}, 7340 (1998), {\it ibid.} {\bf D 60}, 049901 (1999).
\bibitem{gasp1}
M. Gasperini and G. Veneziano, Astropart. Phys. {\bf 1}, 317 (1993);
Mod. Phys. Lett. {\bf A 8}, 3701 (1993); Phys. Rev. {\bf D 50}, 2519
(1994).
\bibitem{vene1}
G. Veneziano, Phys. Lett. {\bf B 265}, 287 (1991).
\bibitem{east2}
R. Easter, K. Maeda and D. Wands, Phys. Rev.  {\bf D 53}, 4247 (1996);
N. Kaloper, R. Madden and K. A. Olive, Nucl. Phys. {\bf B 452}, 677
(1995); Phys. Lett. {\bf B 371}, 34 (1996).
\bibitem{anton1}
I. Antoniadis, J. Rizos and K. Tamvakis, Nucl. Phys. {\bf B 415},
497 (1994).
\bibitem{east1}
R. Easter and K. Maeda, Phys. Rev.  {\bf D 54}, 7252 (1996).
\bibitem{gross}
D. Gross and J. H. Solan, Nucl. Phys. {\bf B 291}, 41 (1987).
\bibitem{anton2}
I. Antoniadis, E. Gava and K. S. Narain, Nucl. Phys. {\bf B 383},  93
(1992).
\bibitem{anton3}
I. Antoniadis, E. Gava and K. S. Narain, Phys. Lett. {\bf B 283},
209 (1992).
\bibitem{rizos1}
J. Rizos and K. Tamvakis, Phys. Lett. {\bf B 326}, 57 (1994).
\bibitem{erd}
A. Erd$\acute{e}$lyi, W. Magnus, F. Oberhettinger and F. G.  Tricomi,
{\it Higher Transcendental Functions}, (MacGraw-hill, New York 1955),
Vol. 3.
\bibitem{kawai2}
S. Kawai and J. Soda,
Phys.Rev. {\bf D59}, 063504 (1999).
\bibitem{kawai1}
S. Kawai, M. Sakagami and J. Soda,
Phys.Lett. {\bf B437}, 284-290 (1998).
\bibitem{soda1}
J. Soda, M. Sakagami and S. Kawai,
in Proceedings "{\it Current Topics in Mathematical Cosmology}" of {\em the International Seminar on Mathematical Cosmology}, edited by
M. Rainer and  H.-J. Schmidt (World Scientific PC Singapore, 1998) , p. 302-309, (gr-qc/9807056).
\bibitem{kawai3}
S. Kawai, M. Sakagami and J. Soda,
in Proceedings of {\em the Seventh Workshop on General Relativity and Gravitation}, edited by Y. Eriguchi et.al., pp.249, (gr-qc/9901065).
\bibitem{king}
D. H. King, Phys. Rev.  {\bf D 44}, 2356 (1991).

\end{thebibliography}
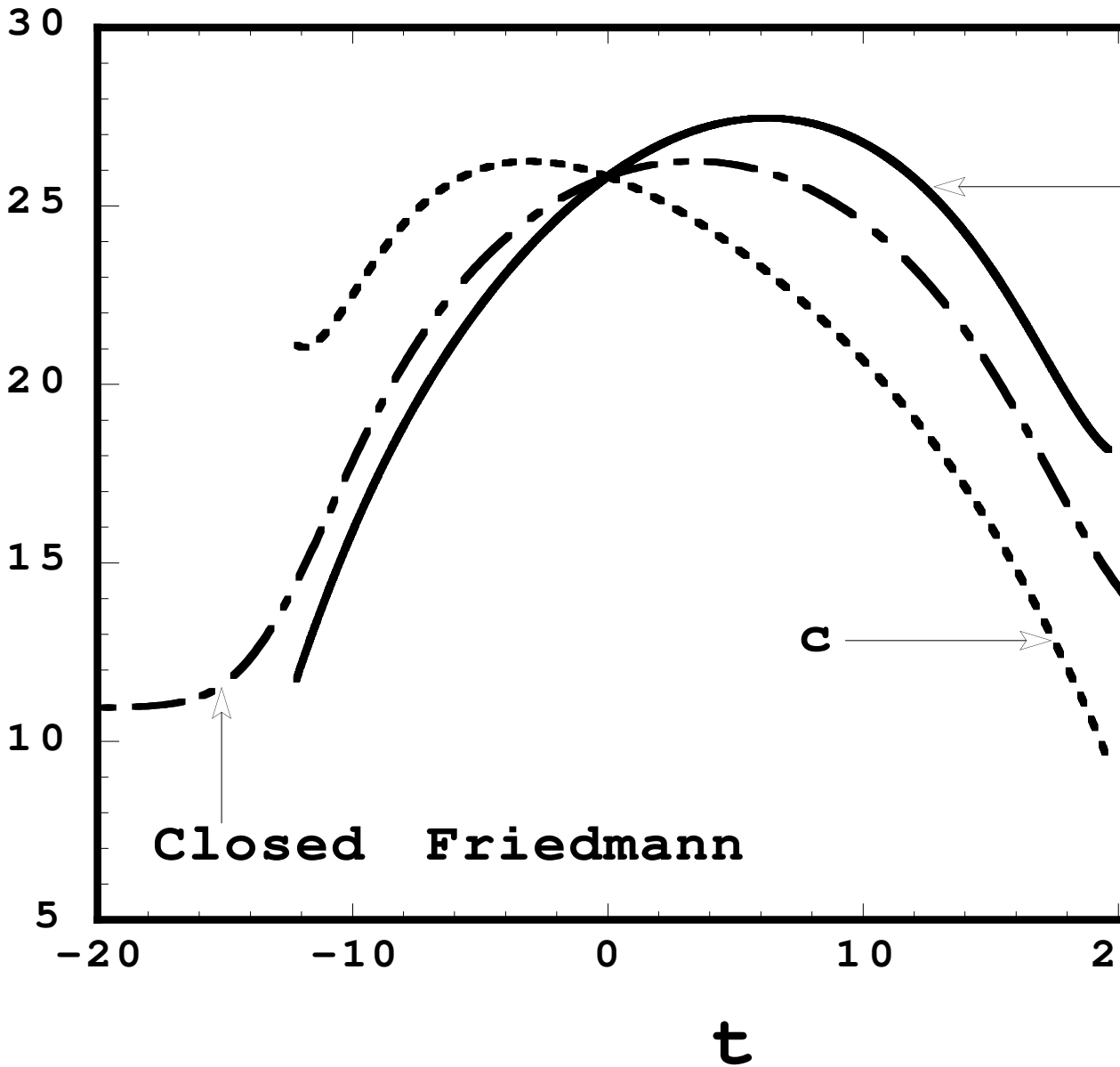
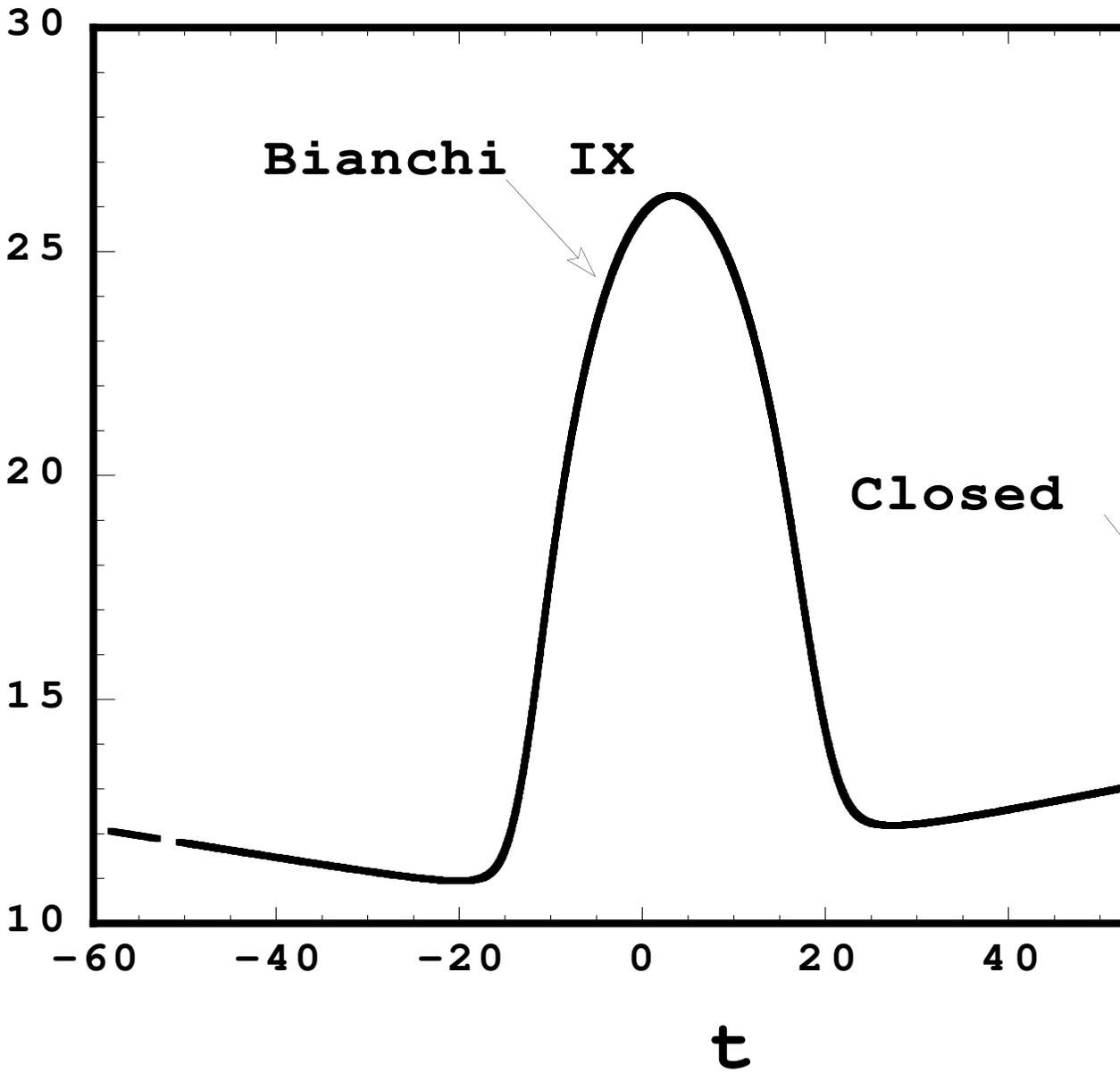
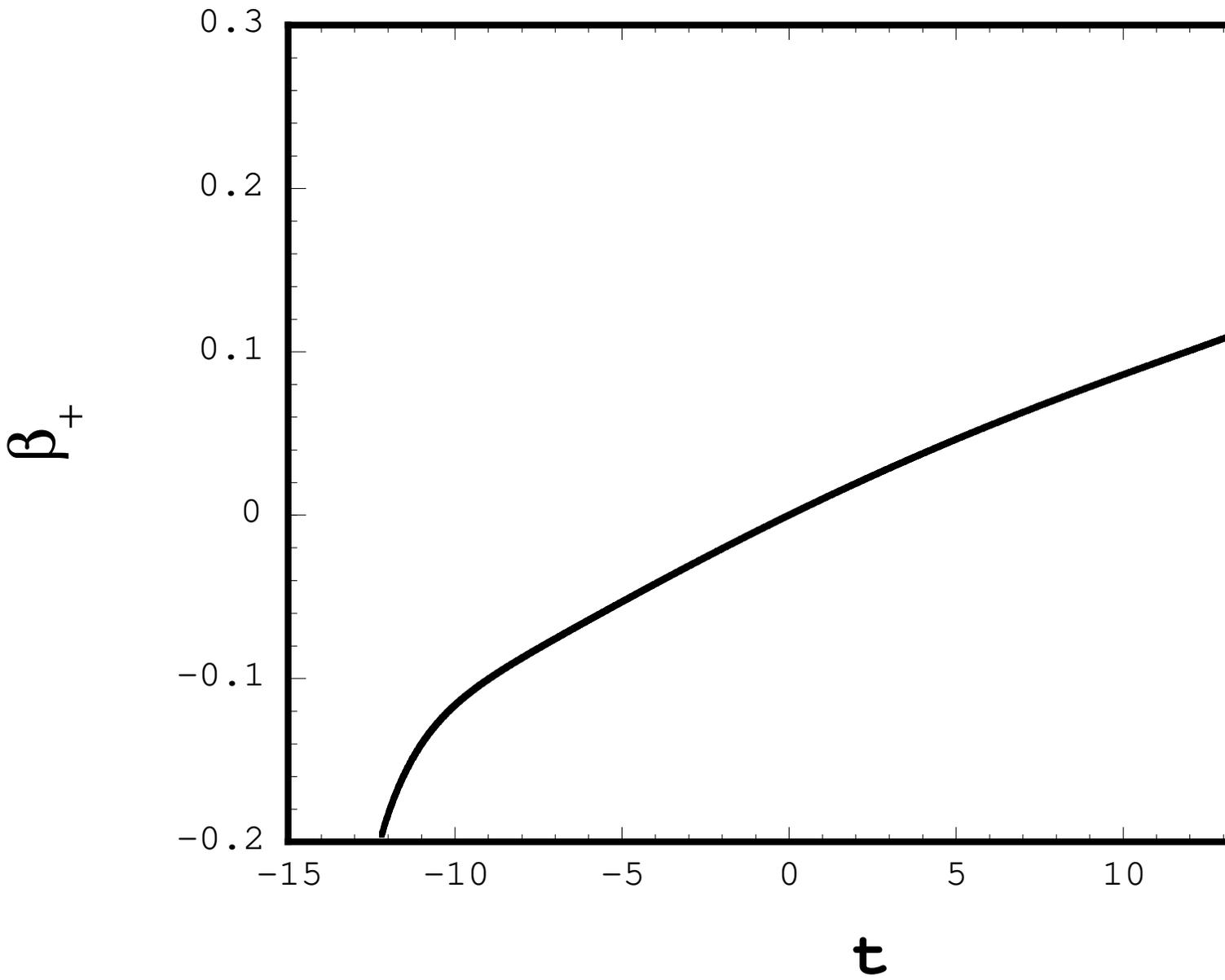
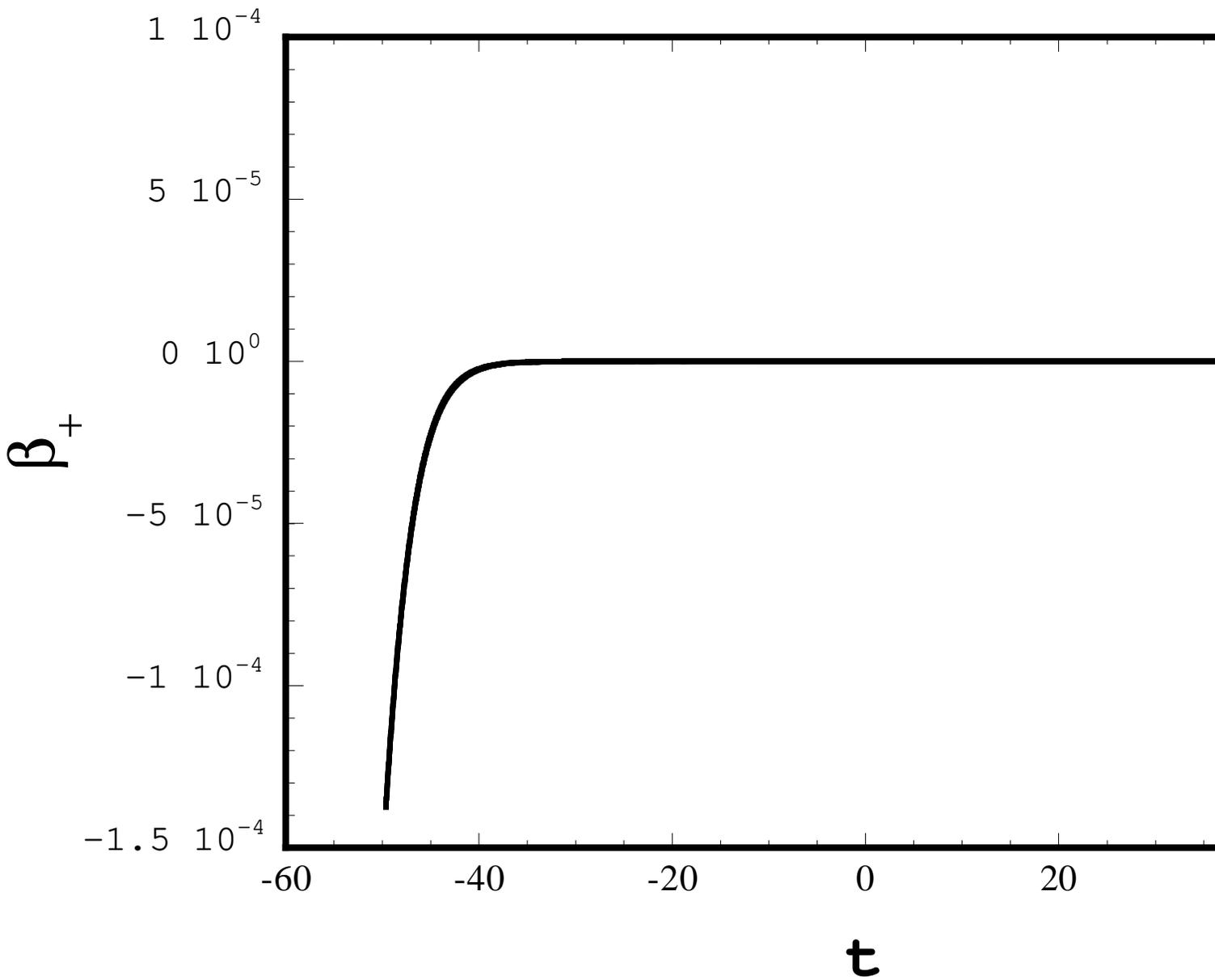
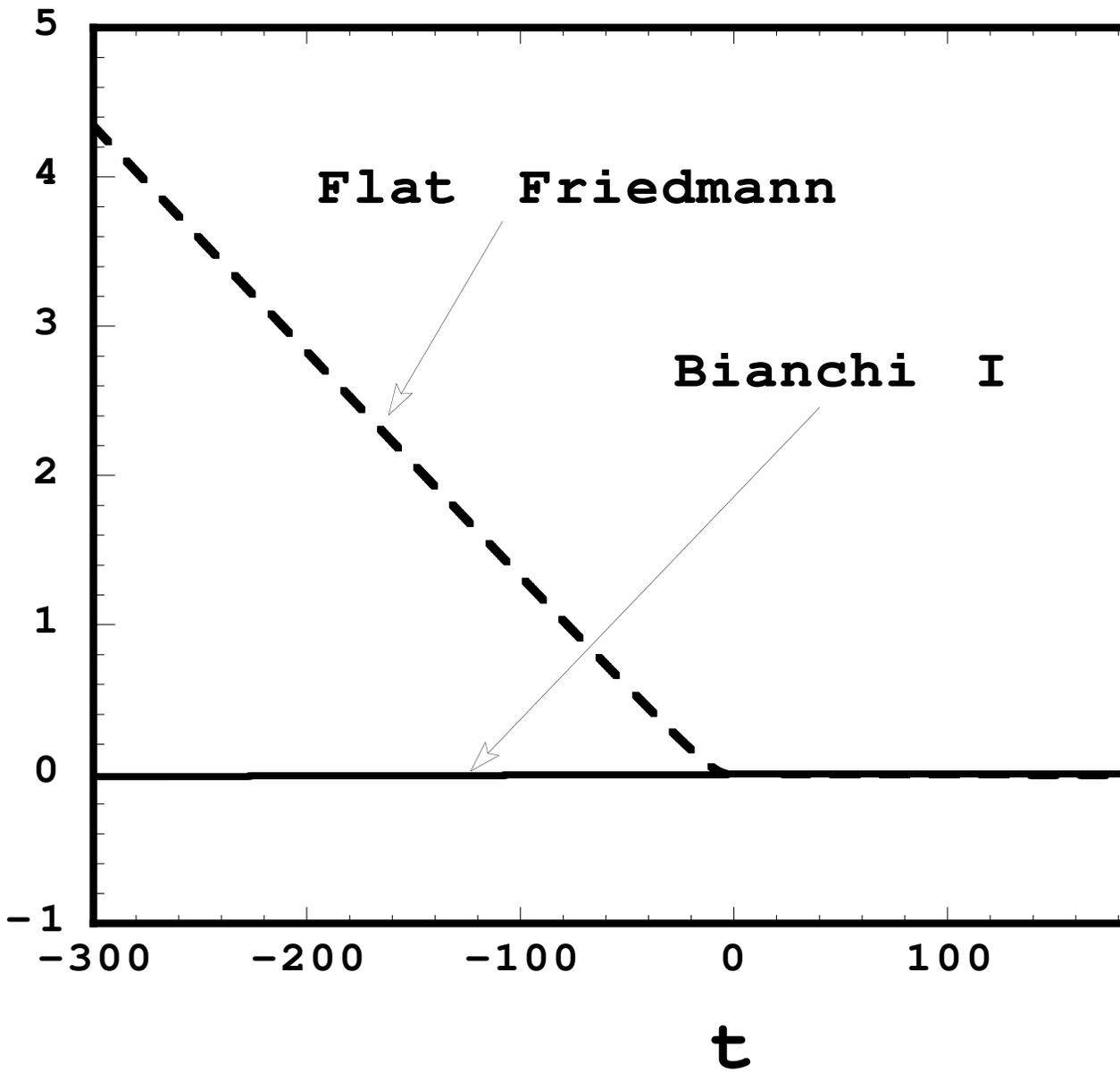
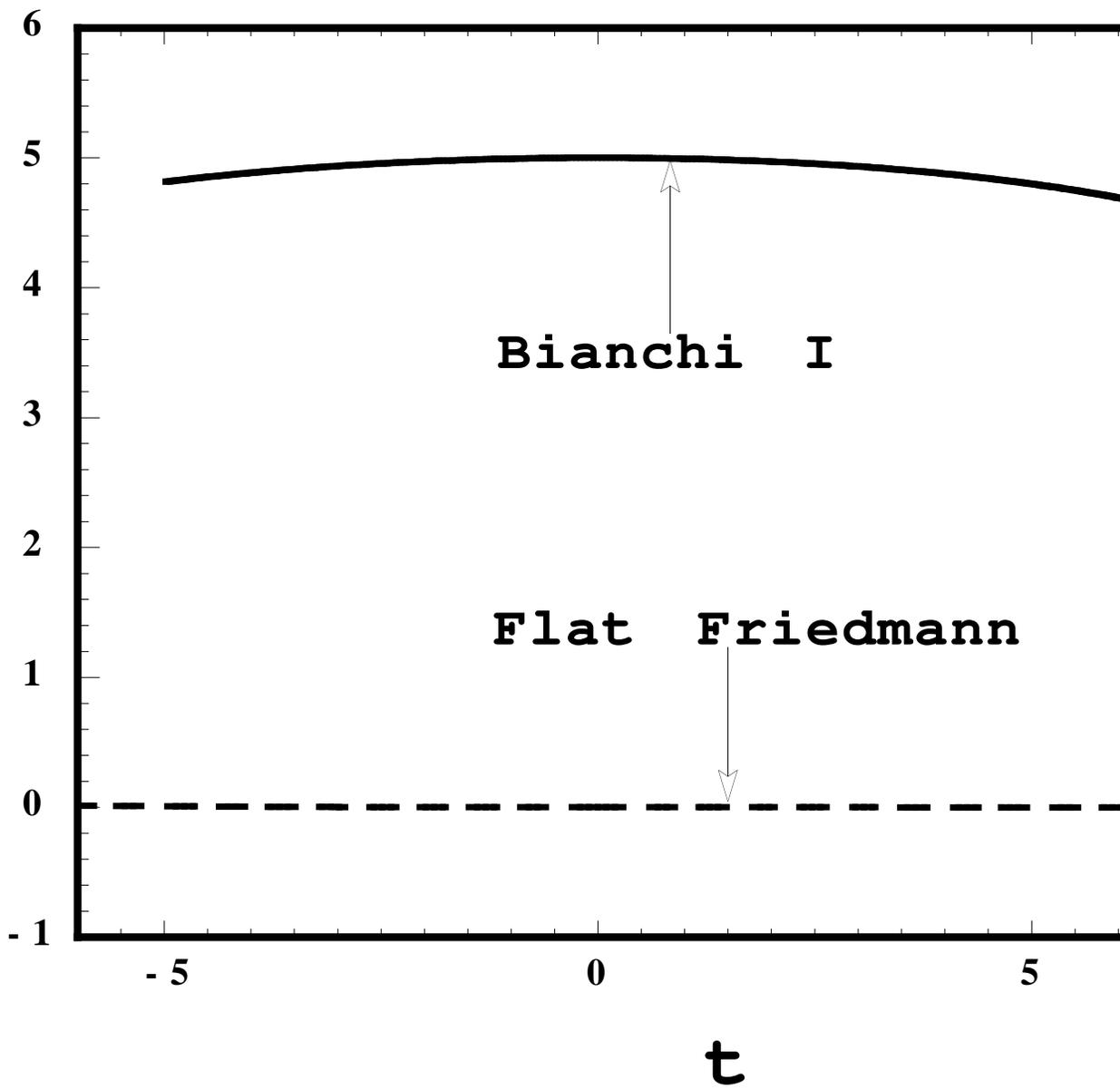
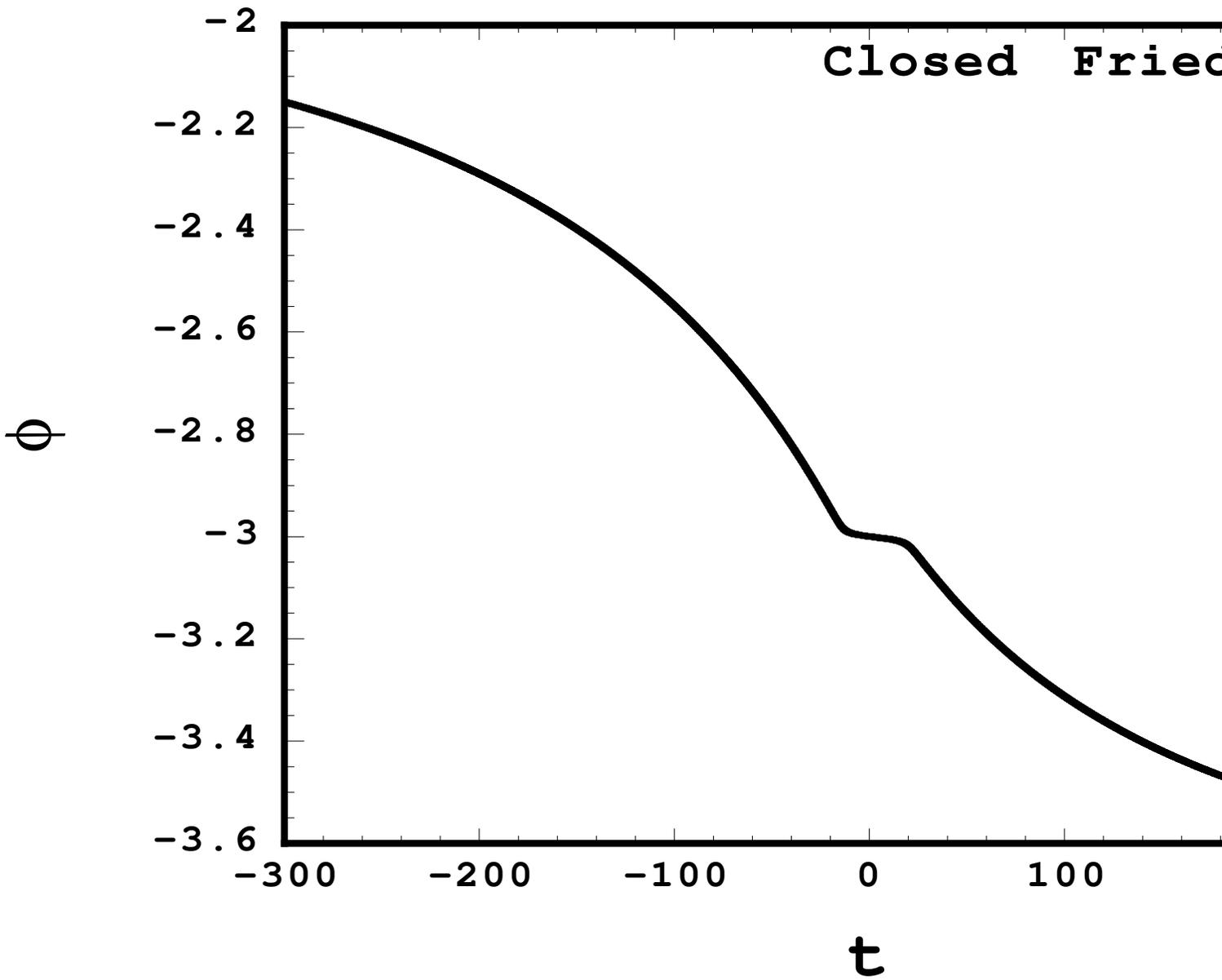
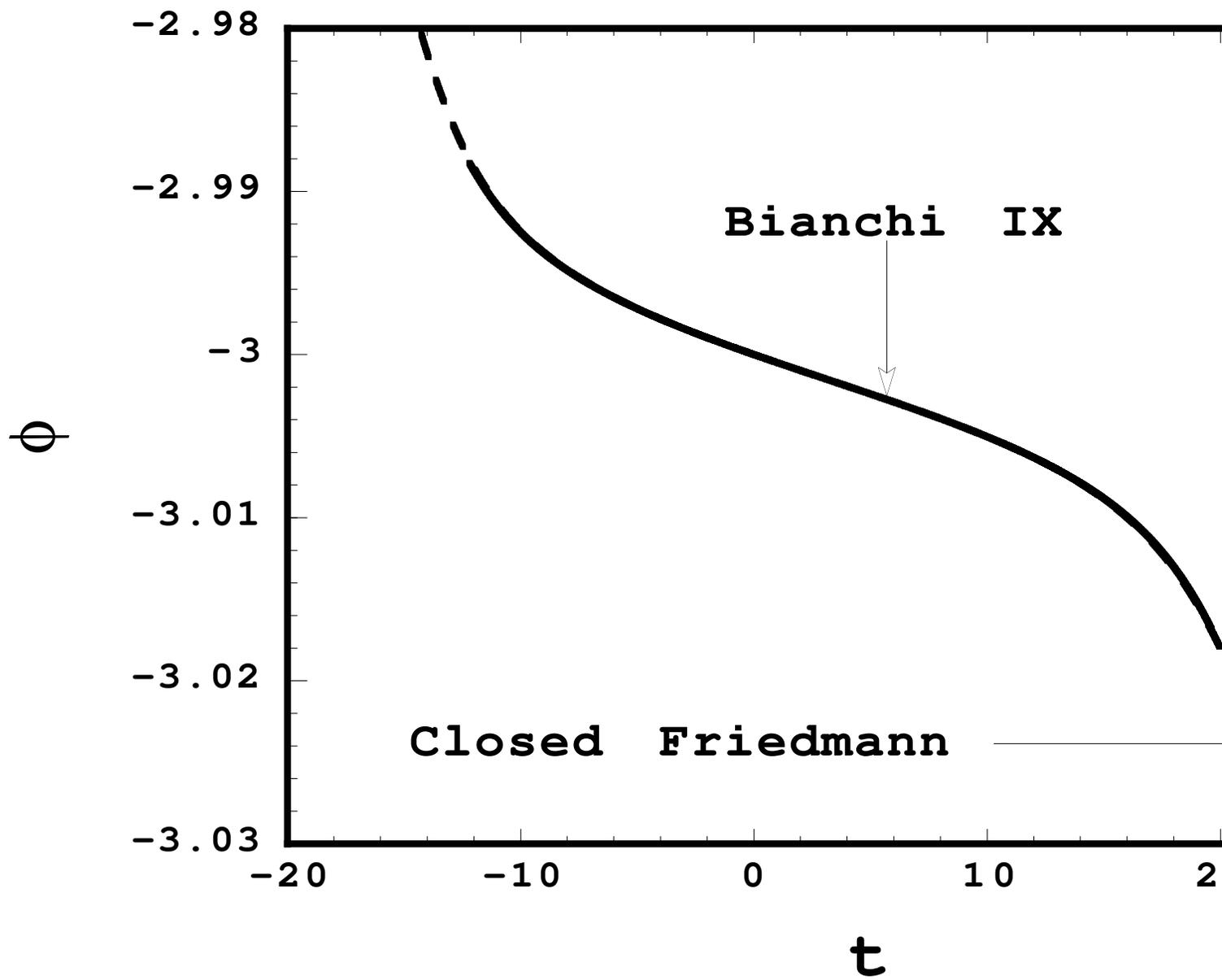
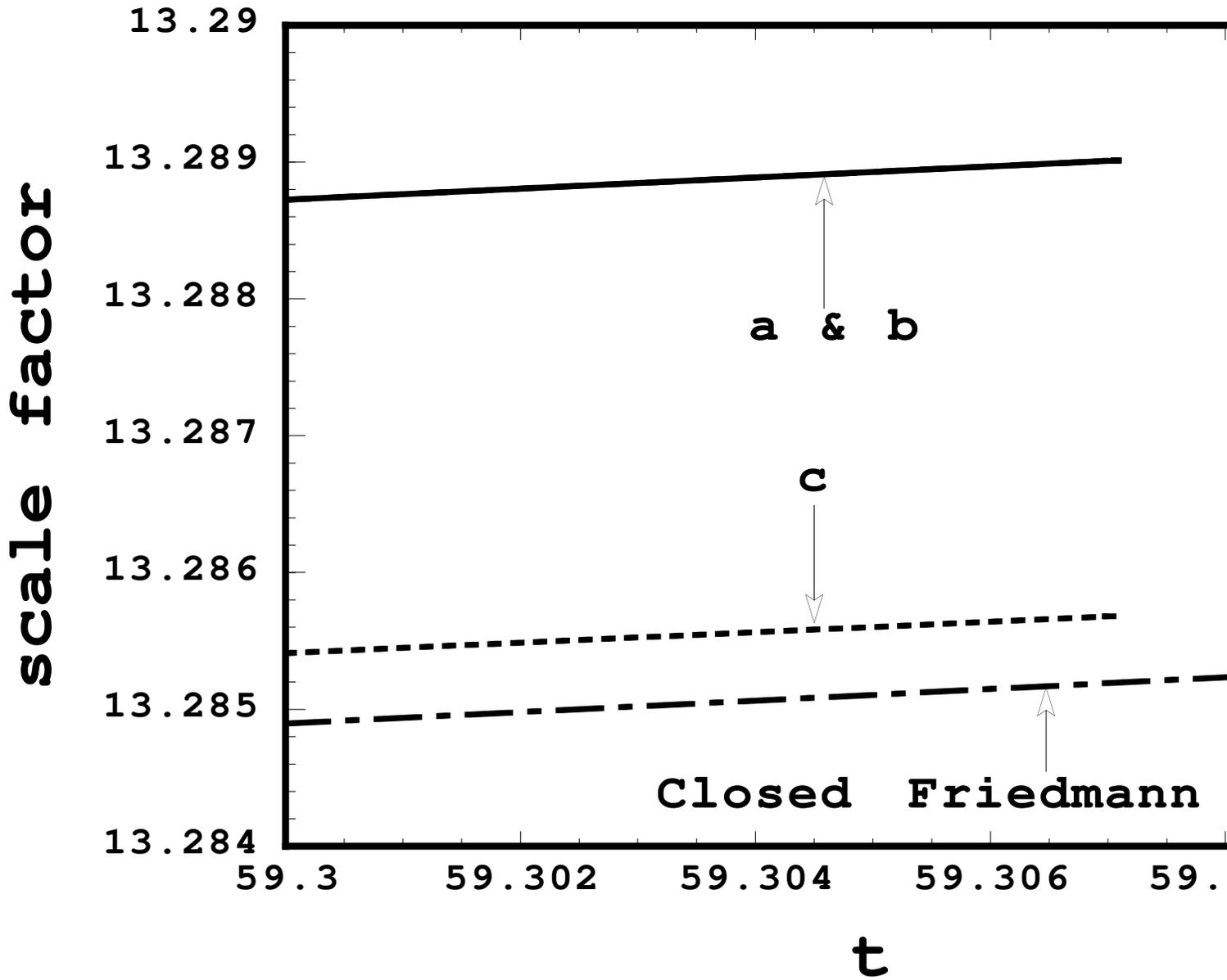
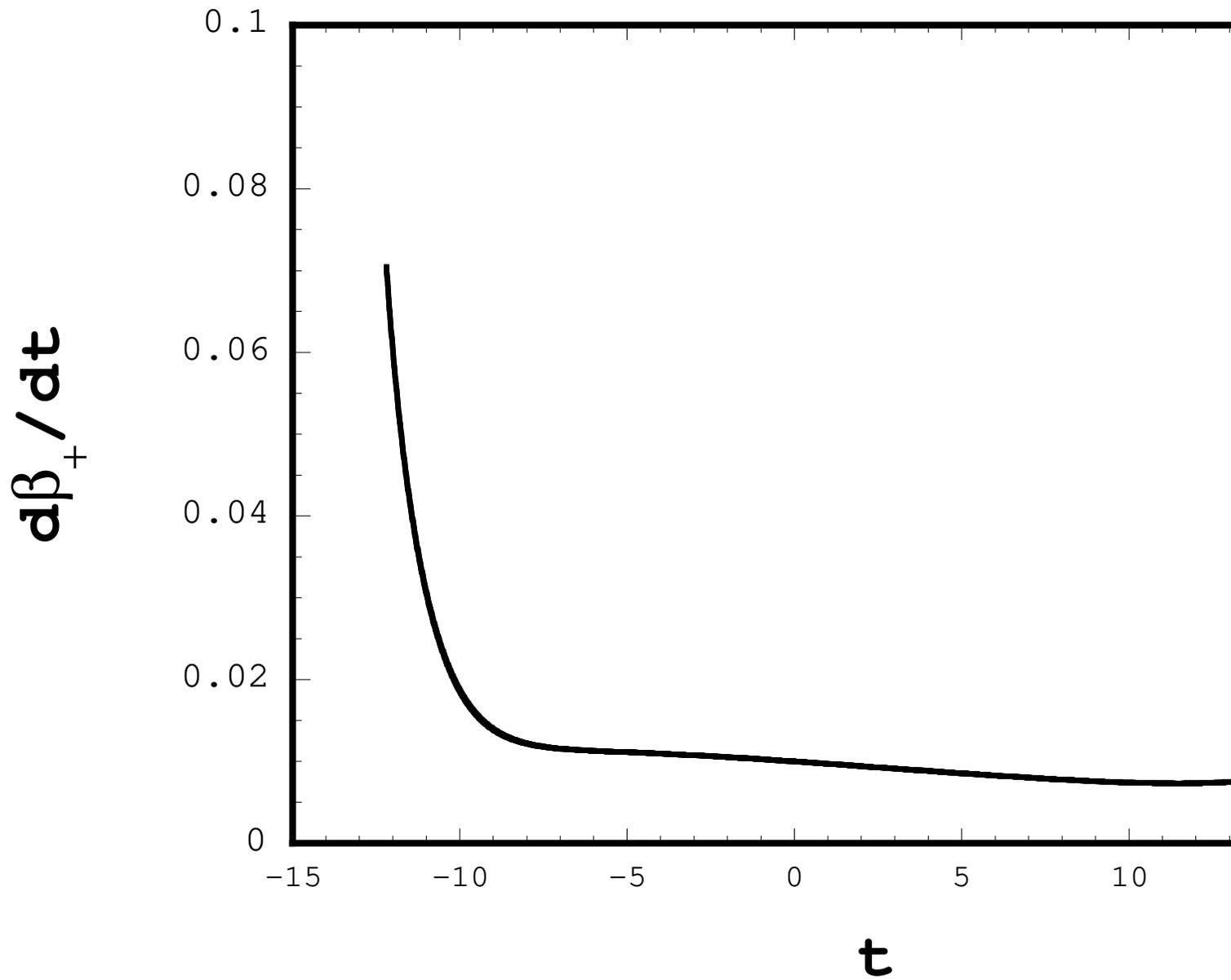
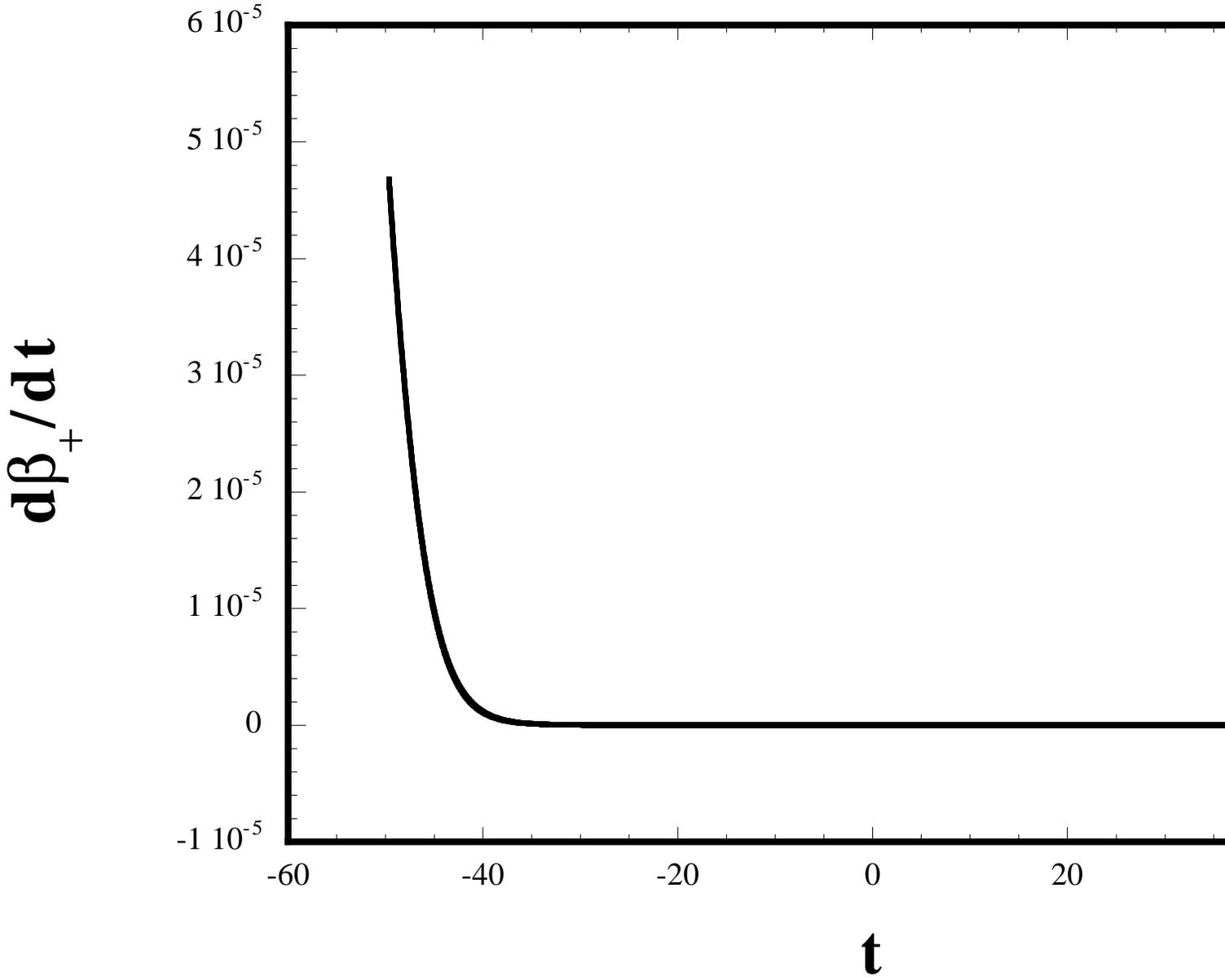
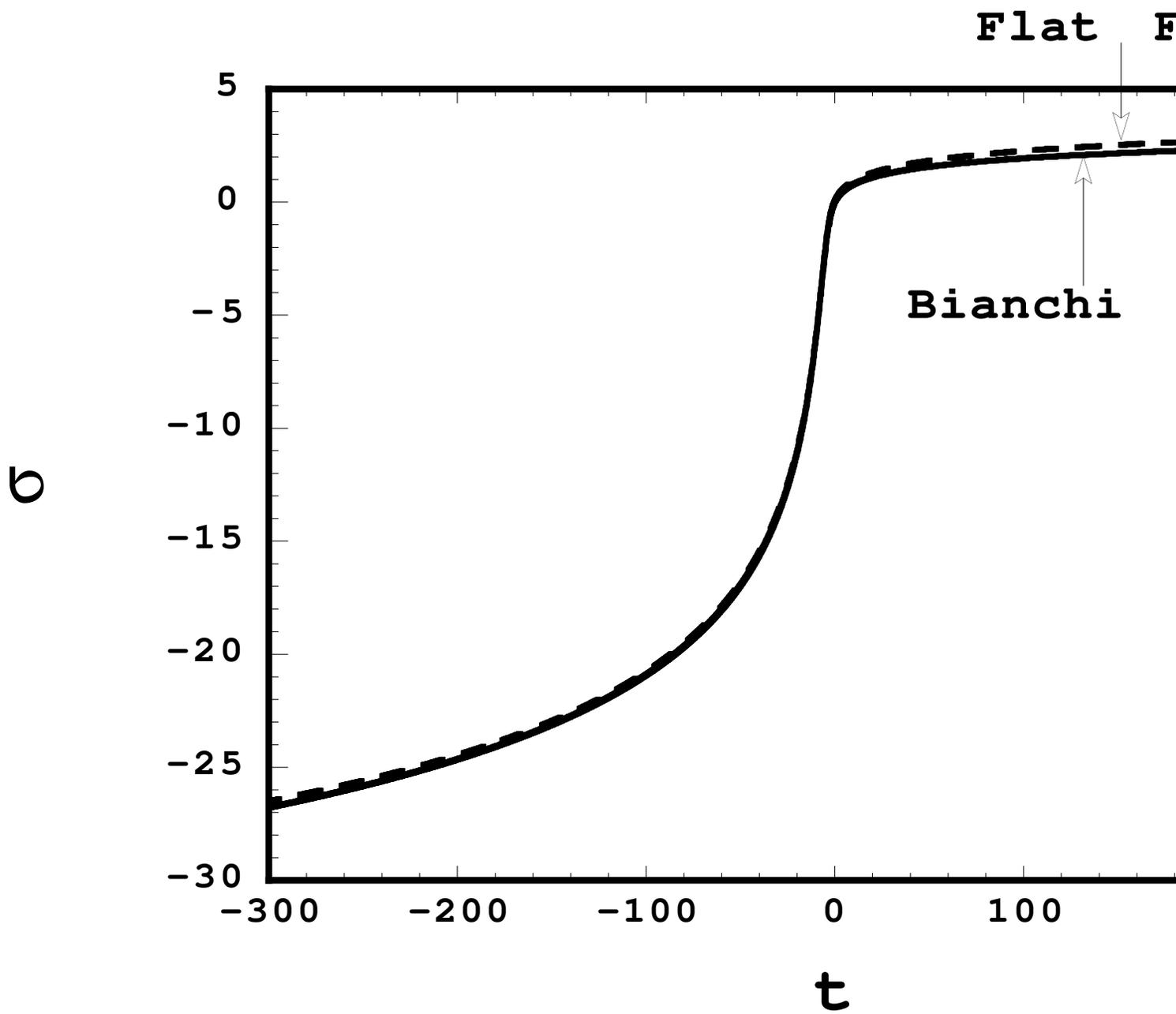
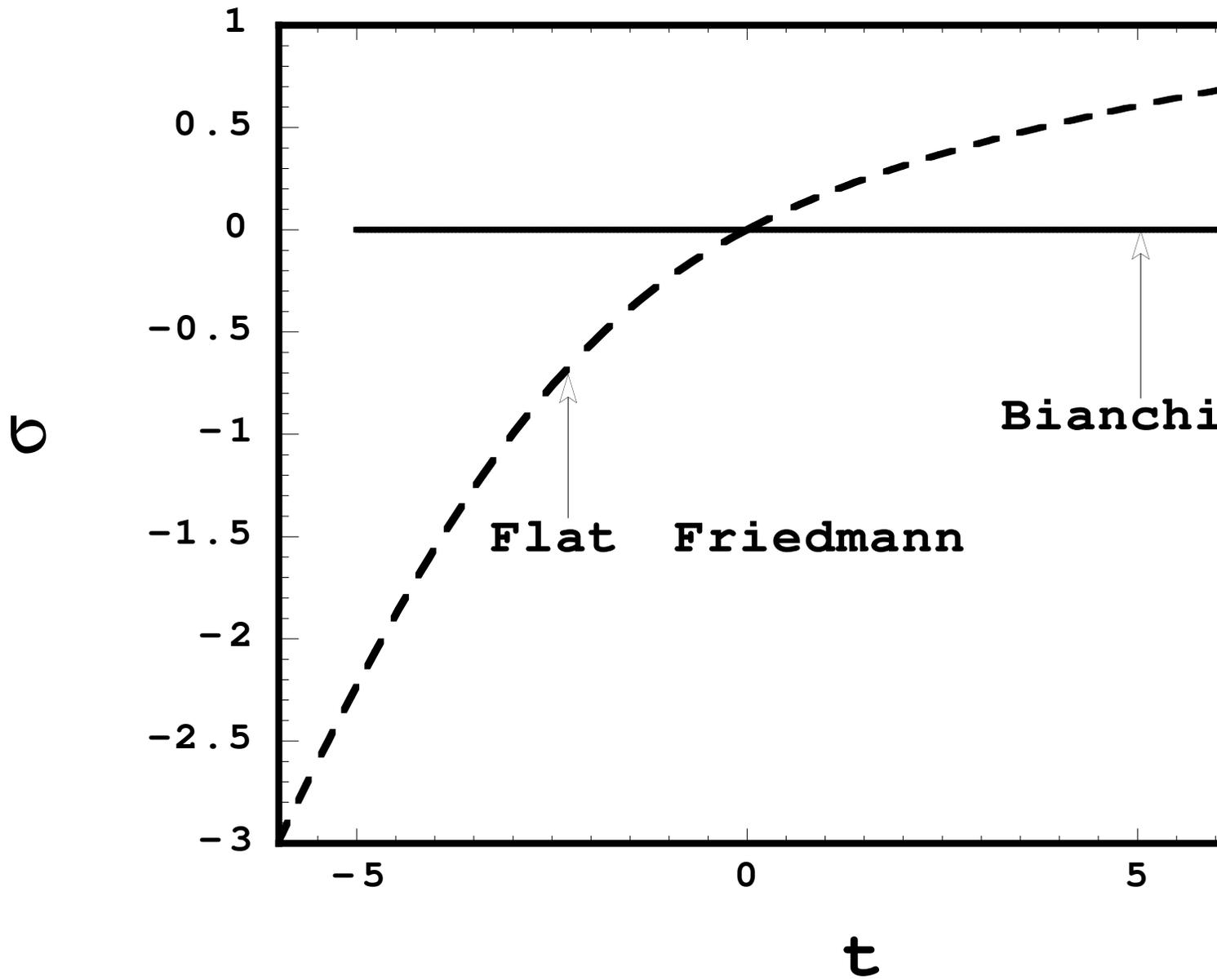
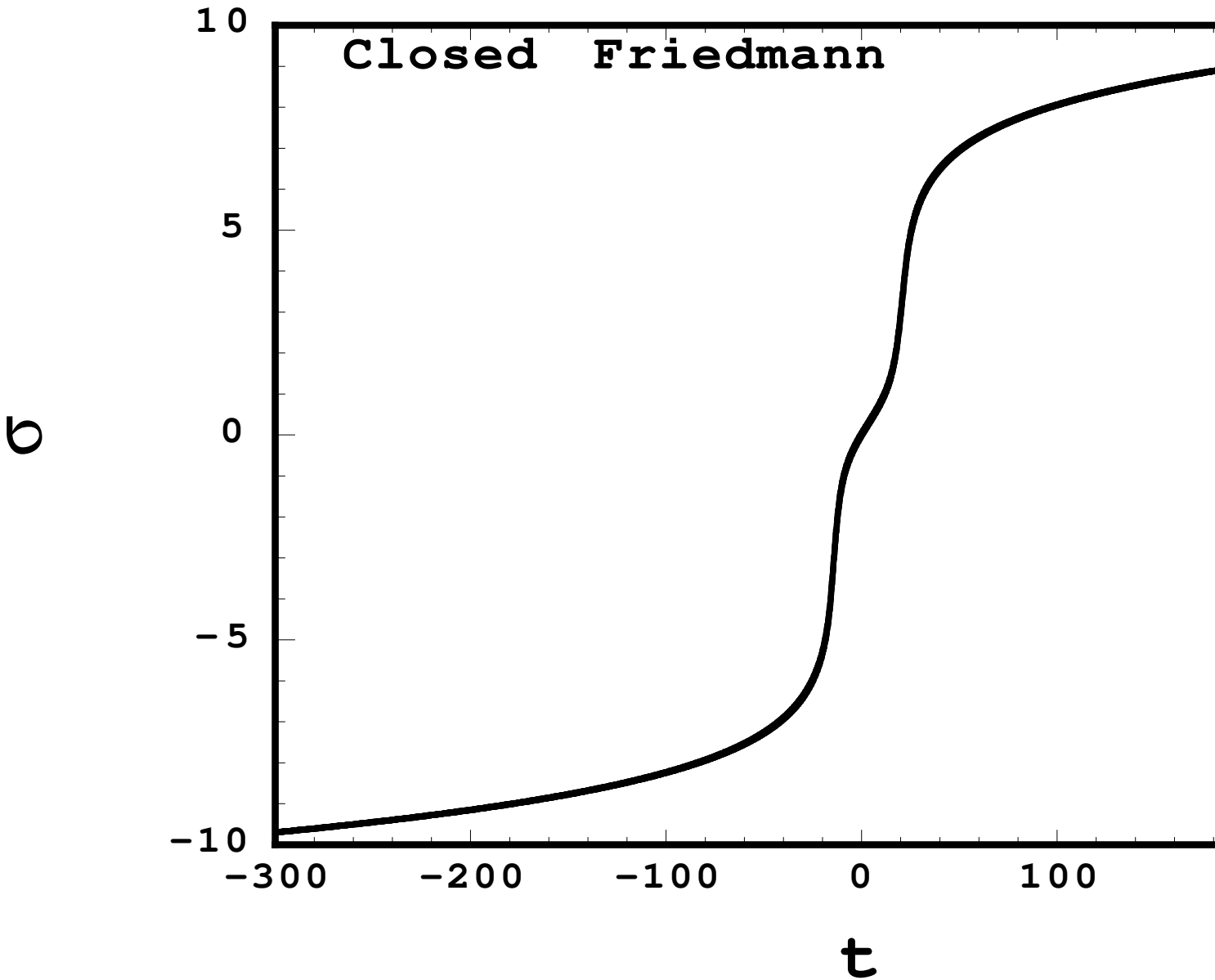
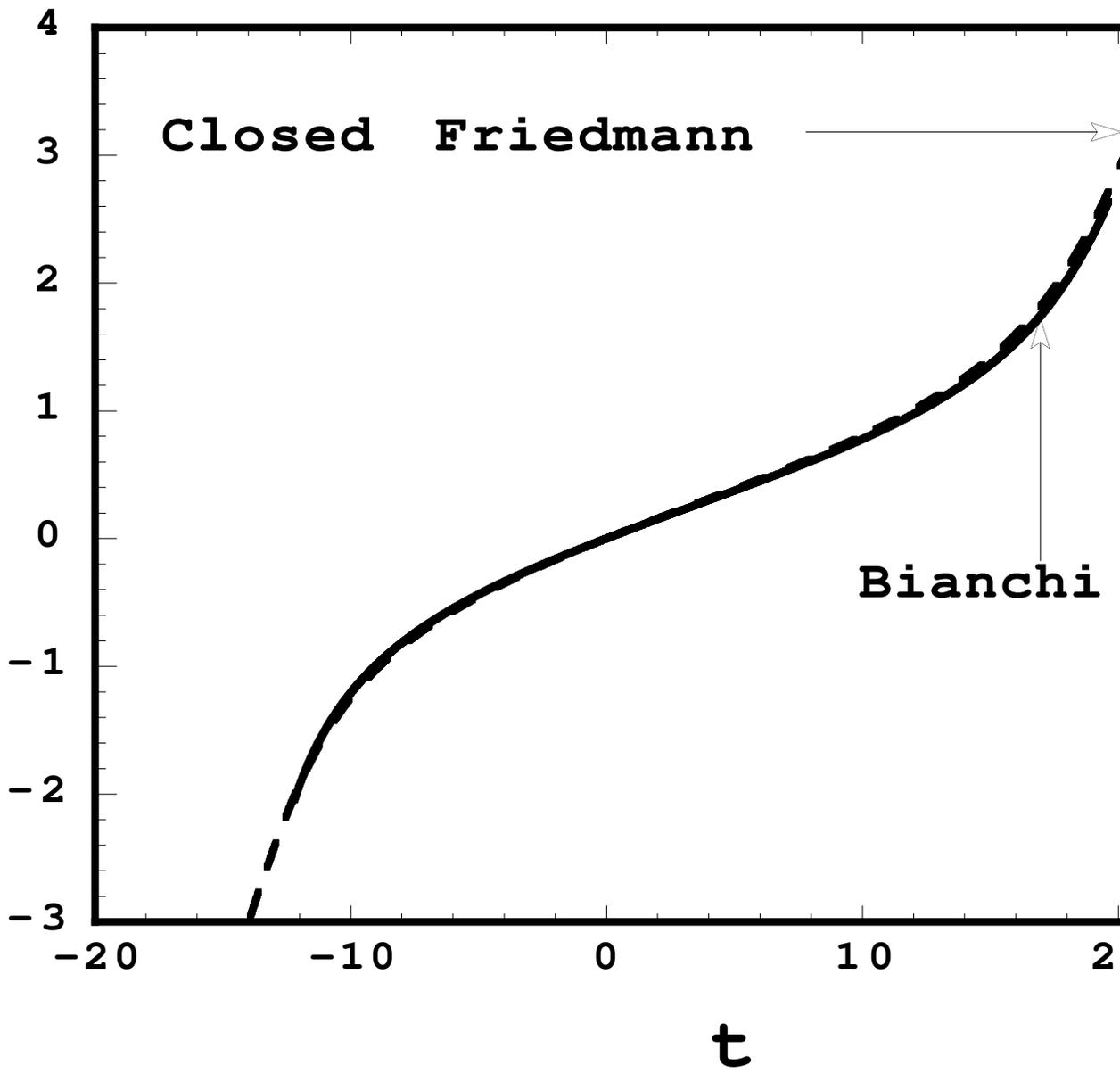
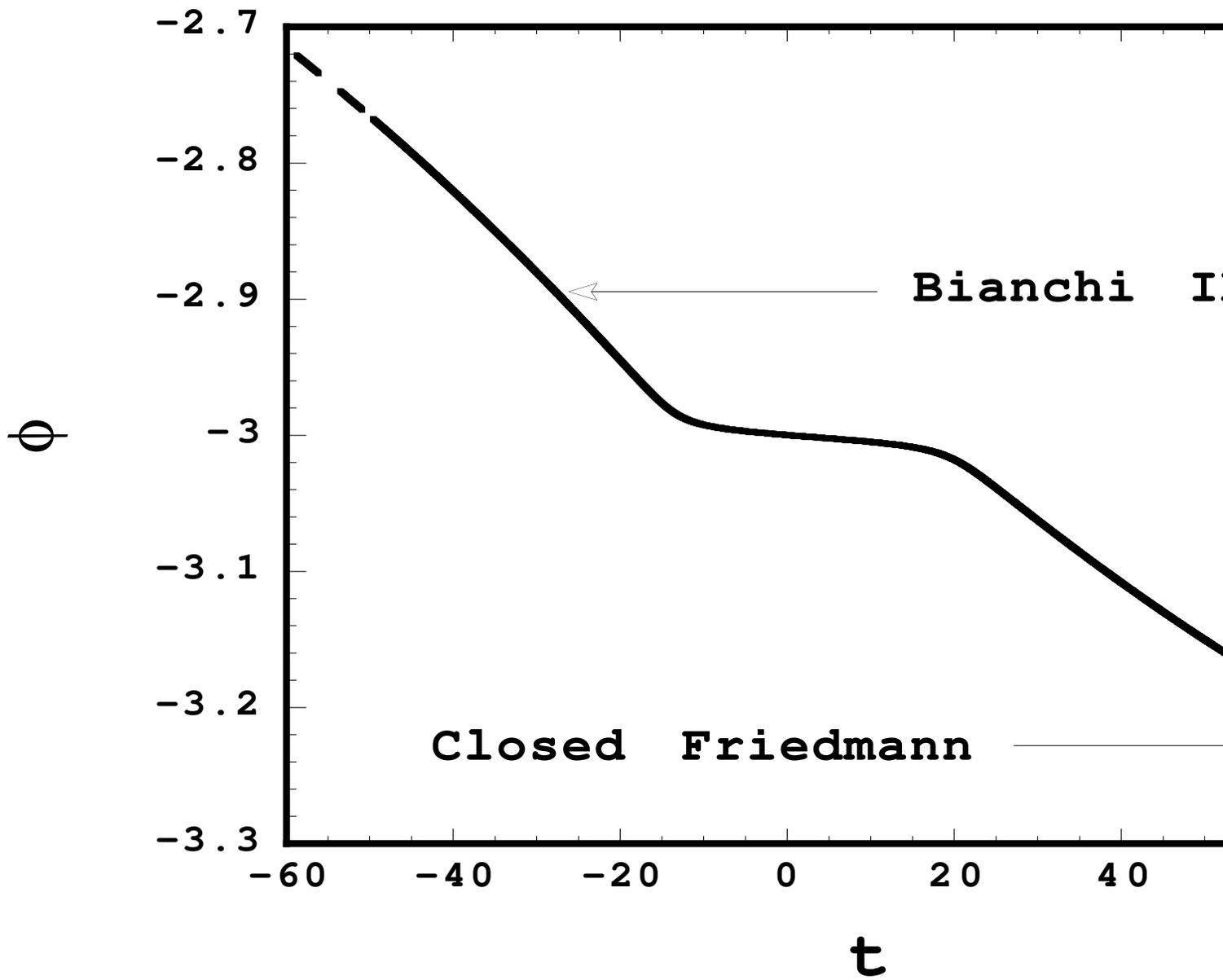
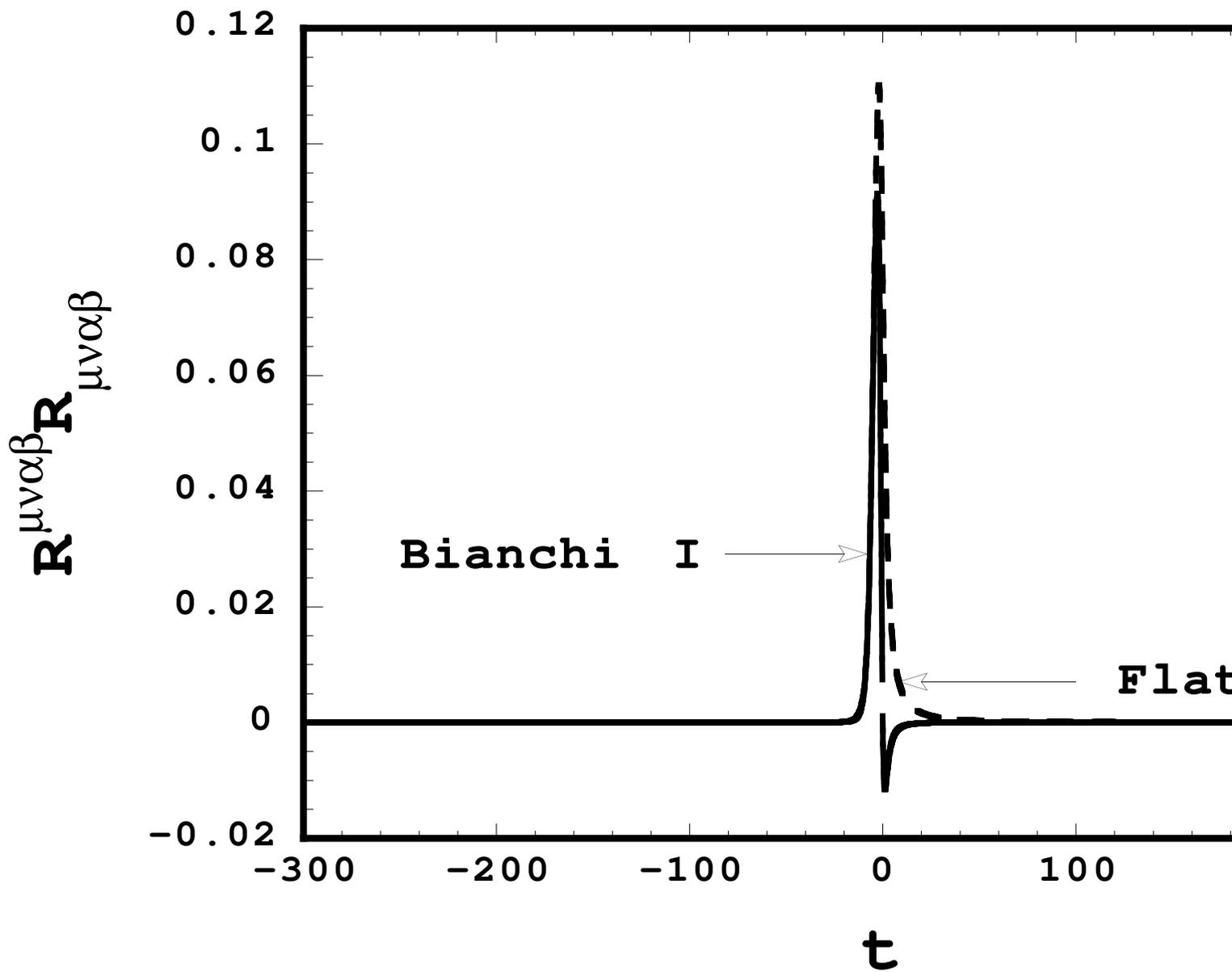
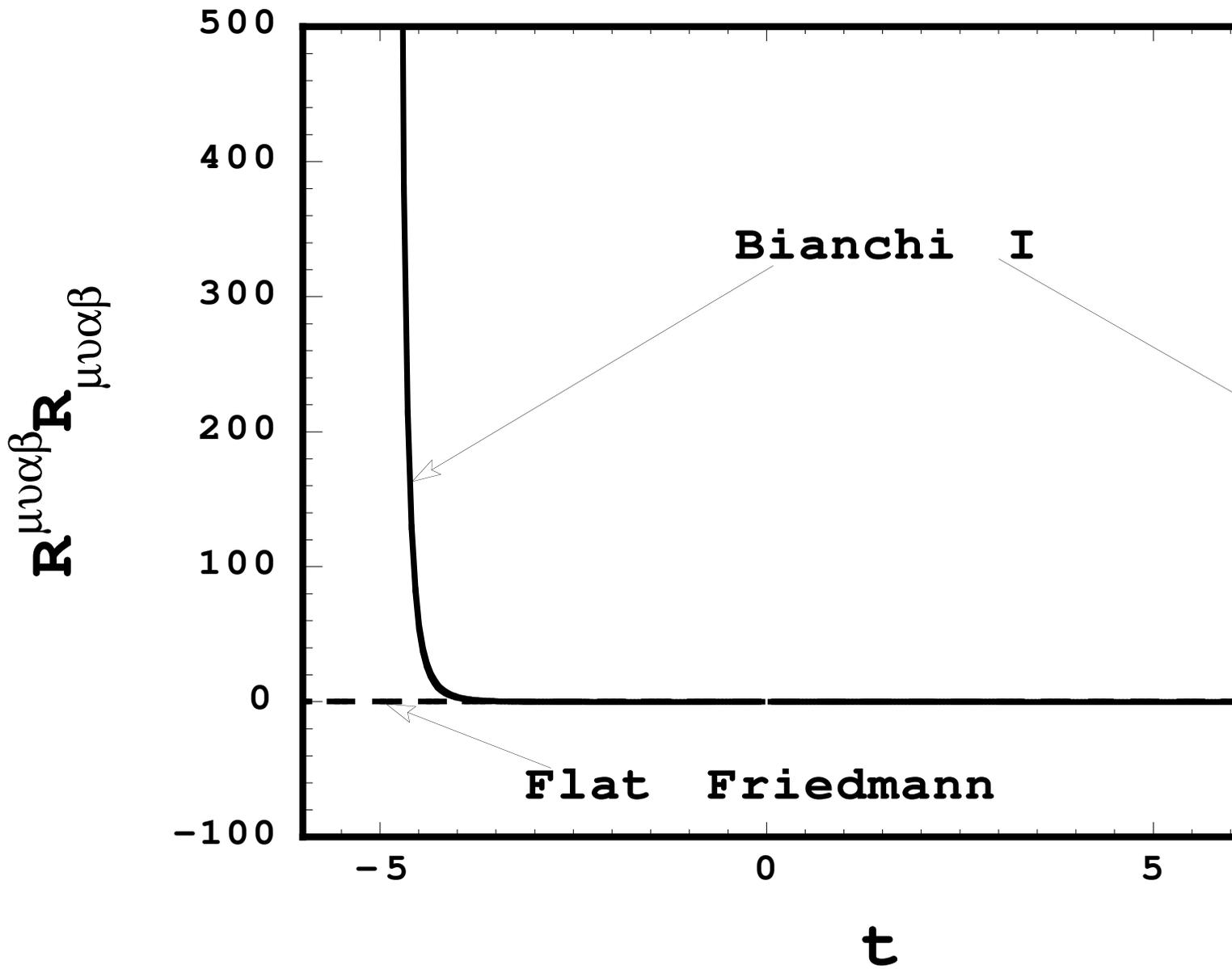
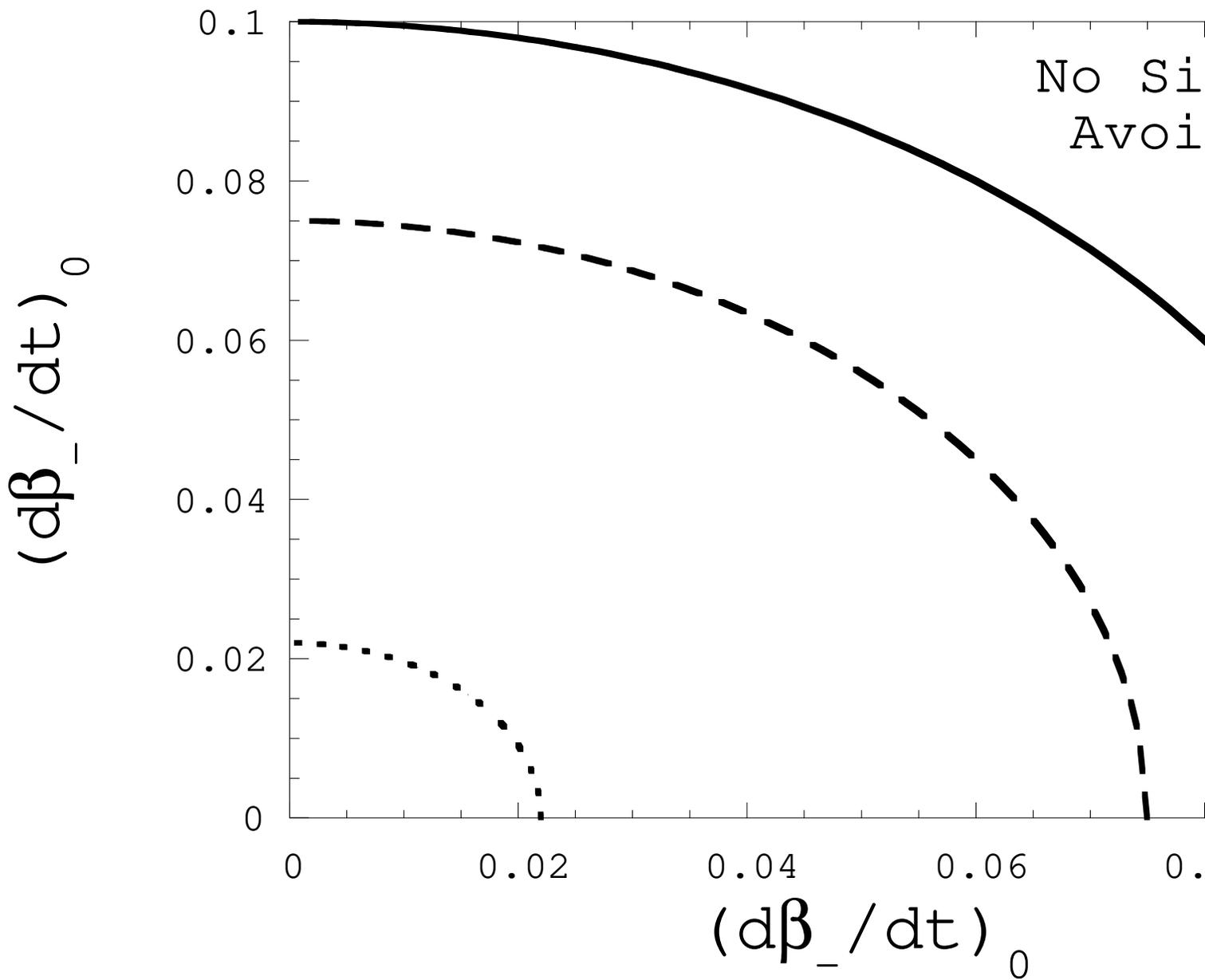
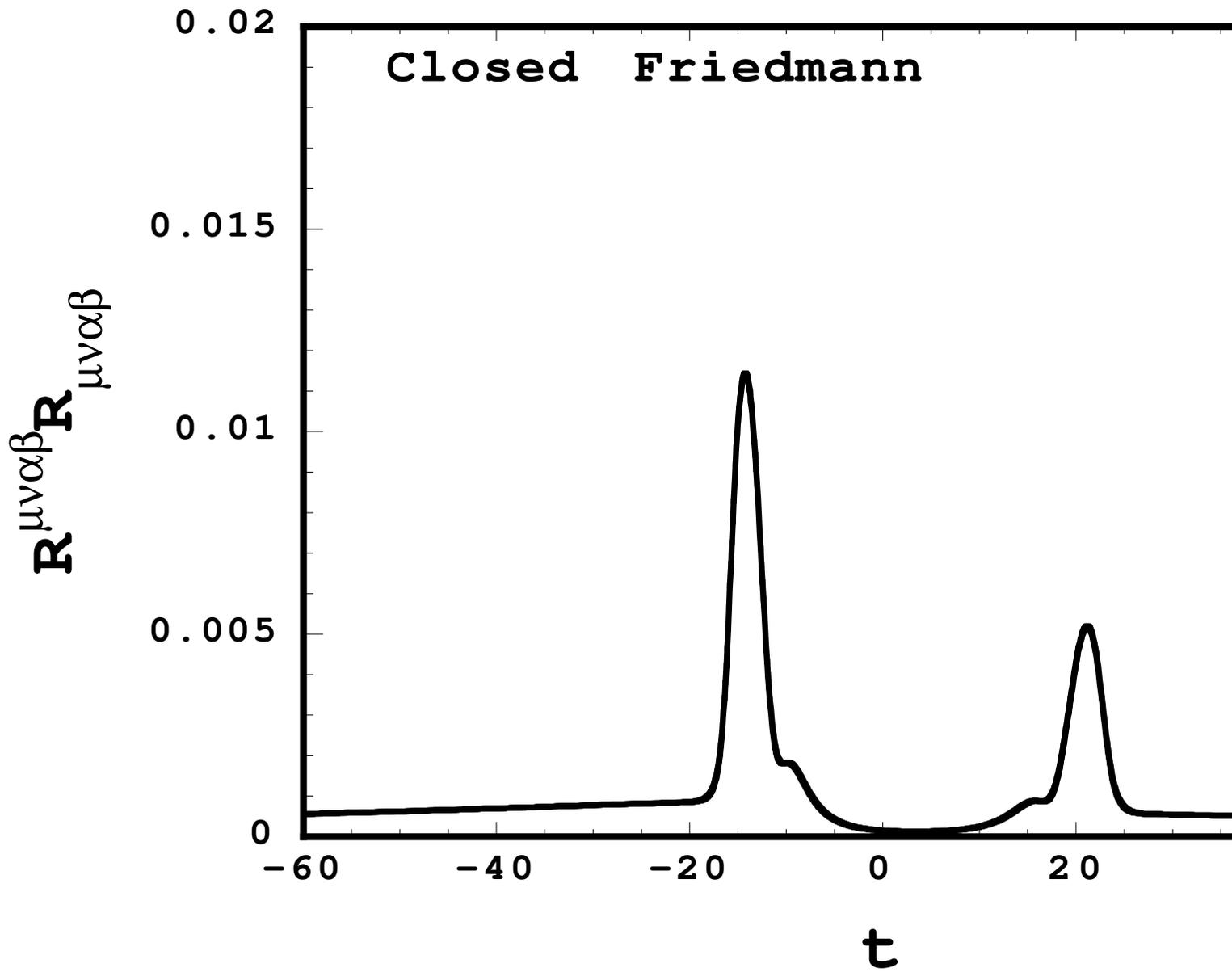
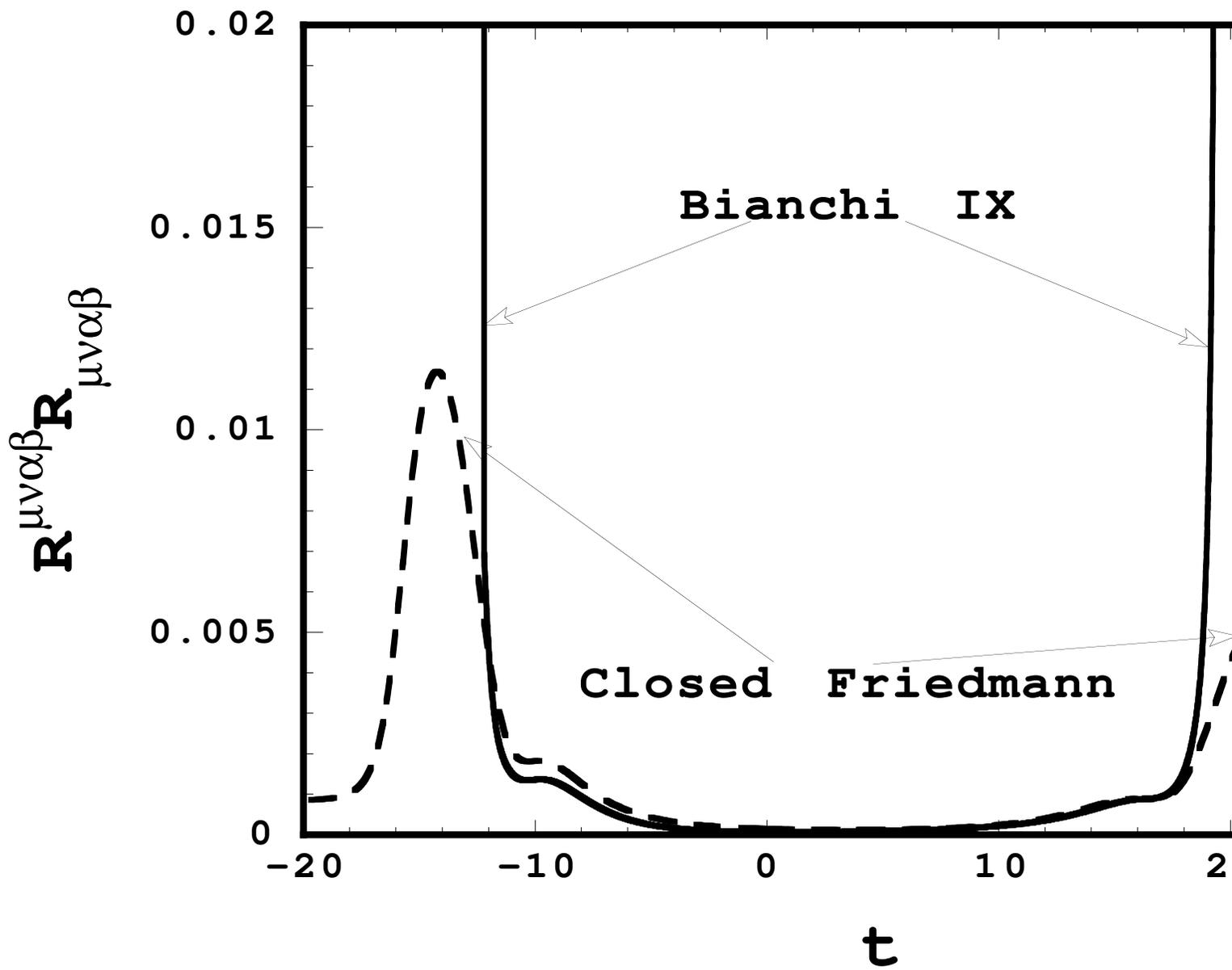
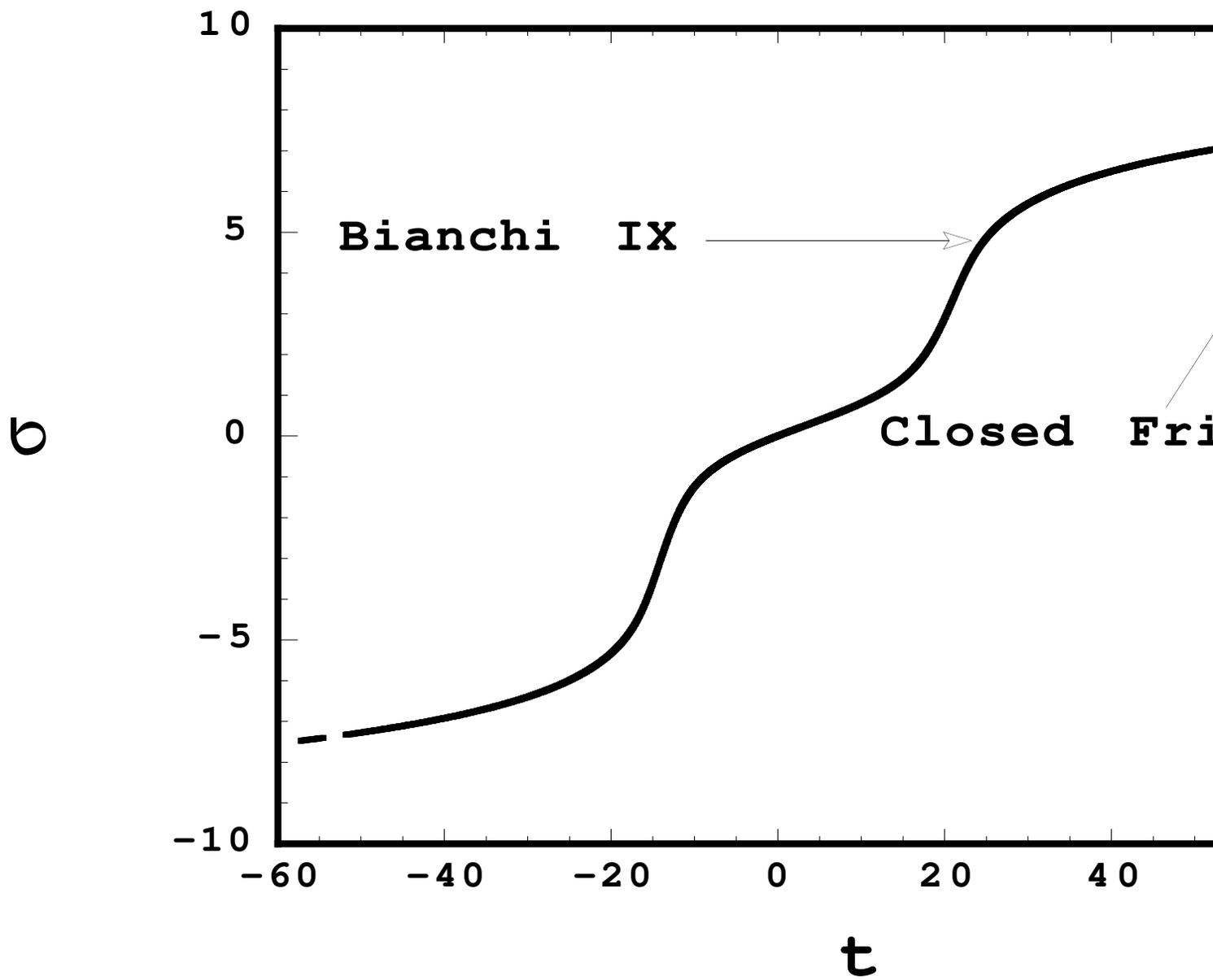
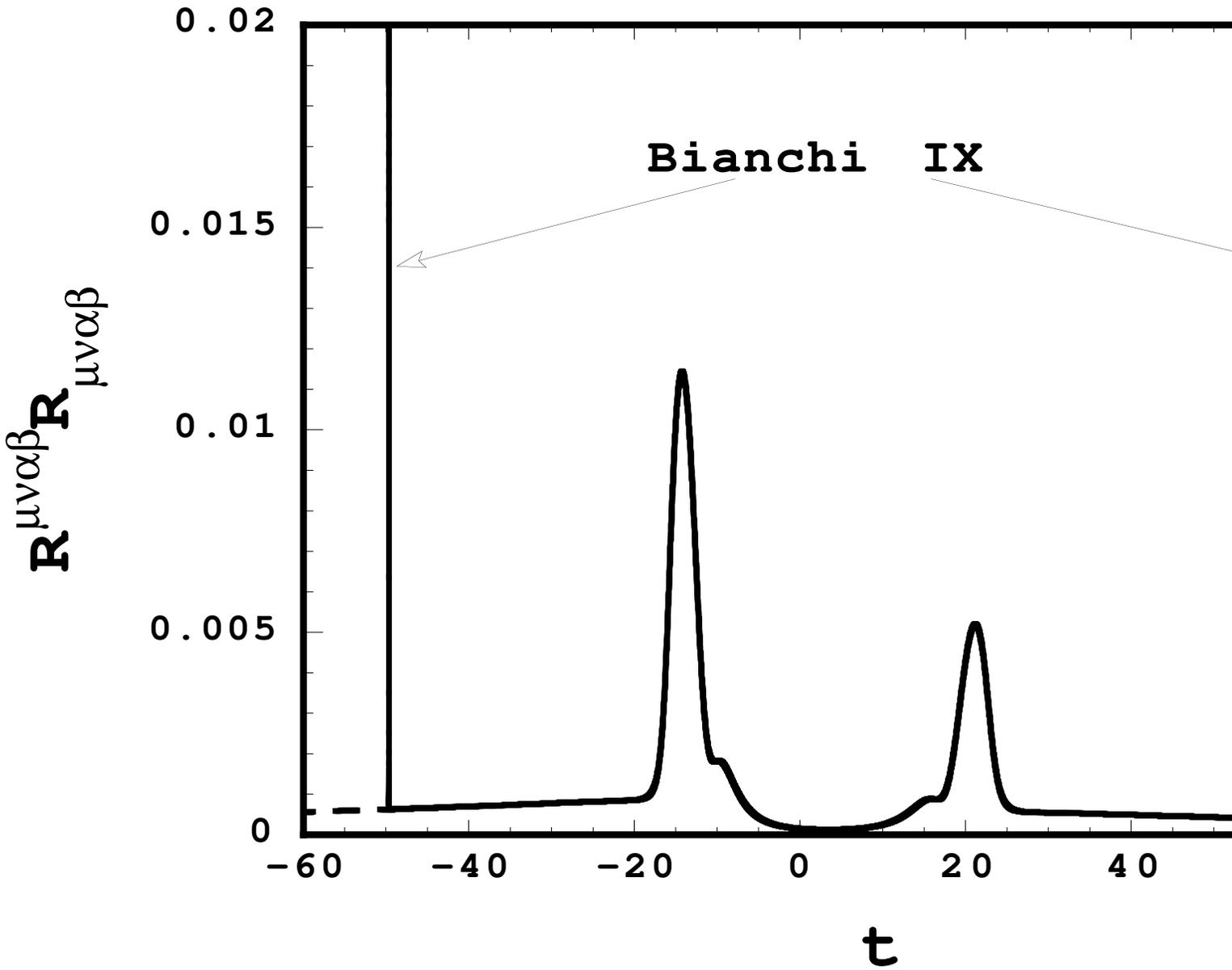
\end{document}